\documentclass[nonacm, sigplan]{acmart}

\usepackage{enumitem}
\usepackage[capitalise]{cleveref}

\usepackage{hyperref}   % Load hyperref first
\usepackage{algorithmic}
\usepackage{graphicx}
\usepackage{subcaption} 
\usepackage[linesnumbered,ruled,vlined]{algorithm2e}
\SetKw{require}{\textbf{Require:}}
\SetKw{ensure}{\textbf{Ensure:}}
\SetKwProg{function}{Function}{ do}{}
\SetKwProg{upon}{Upon}{ do}{}

\crefalias{algocf}{algorithm}

\newcommand{\system}{Yoimiya\xspace}
\newcommand{\zk}{ZK-SNARK\xspace}
\renewcommand{\paragraph}[1]{\vskip 0.05in \noindent\textbf{#1.}}

\begin{document}

\title{\system: A Scalable Framework for Optimal Resource Utilization in ZK-SNARK Systems}

\author{Zheming Ye}
\affiliation{
  \institution{East China Normal University}
  \city{Shanghai}
  \country{China}
}
\email{zhmye@stu.ecnu.edu.cn}

\author{Xiaodong Qi}
\affiliation{
  \institution{Nanyang Technological University}
  % \city{Shanghai}
  \country{Singapore}
}
\email{xiaodong.qi@ntu.edu.sg}

\author{Zhao Zhang}
\affiliation{
  \institution{East China Normal University}
  \city{Shanghai}
  \country{China}
}
\email{zhzhang@dase.ecnu.edu.cn}

\author{Cheqing Jin}
\affiliation{
  \institution{East China Normal University}
  \city{Shanghai}
  \country{China}
}
\email{cqjin@dase.ecnu.edu.cn}

%\author{...} % removed for anonymity

% Zero-knowledge proofs (ZKPs) are increasingly employed in privacy-preserving technologies, blockchain applications, and secure authentication. However, several challenges remain in their current implementation. The high memory footprint not only limits the scalability of ZKP systems but also hinders their applicability on resource-constrained platforms. Furthermore, in most implementations, the witness generation and proof computation stages are tightly coupled. While recent optimizations have focused on parallelizing the proof phase, this coupling has caused witness generation to become a bottleneck, leading to inefficiencies in resource utilization and prolonged proof generation time.

% We present \textbf{Yoimiya}, a flexible ZKP framework that addresses these challenges. It introduces an automatic circuit partitioning mechanism that allows users to split ZKP circuits into sequential sub-circuits, reducing memory consumption during proof generation. Furthermore, Yoimiya decouples the witness generation and proof computation stages, organizing them into a pipeline with configurable parallelism. This design improves resource utilization and speeds up proof generation, particularly when generating multiple proofs. By combining partitioning with pipelining, Yoimiya efficiently manages memory in parallelized environments. Extensive experiments demonstrate its effectiveness in improving proof generation efficiency and reducing memory overhead.
\begin{abstract}
%Zero-Knowledge Proof systems, particularly {\zk}s, are widely used in verifiable computation outsourcing applications. Despite the efficiency of ZK-SNARKs in proof verification, generating proofs remains a computationally intensive task, particularly in systems handling continuous streams of proof requests. Existing optimization efforts have primarily focused on accelerating proof computation, often overlooking the inefficiencies in the witness generation phase, which has now become a critical bottleneck.
With the widespread adoption of Zero-Knowledge Proof systems, particularly {\zk}, the efficiency of proof generation, encompassing both the witness generation and proof computation phases, has become a significant concern. While substantial efforts have successfully accelerated proof computation, progress in optimizing witness generation remains limited, which inevitably hampers overall efficiency.
%Existing optimization efforts have primarily focused on accelerating proof computation, often overlooking the inefficiencies in the witness generation phase, which has now become a critical bottleneck.
In this paper, we propose \system, a scalable framework with pipeline, to optimize the efficiency in \zk systems. First,  \system introduces an automatic circuit partitioning algorithm that divides large circuits of {\zk} into smaller subcircuits, the minimal computing units with smaller memory requirement, allowing parallel processing on multiple units. 
%effectively reducing memory consumption while minimizing resource contention.
Second, \system decouples witness generation from proof computation, and achieves simultaneous executions over units from multiple circuits. Moreover, \system enables each phase scalable separately by configuring the resource distribution to make the time costs of the two phases aligned, maximizing the resource utilization. 
%a configurable pipeline framework that decouples witness generation from proof computation, allowing flexible control over parallelism, thereby improving CPU utilization and proof generation speed. However, the pipeline introduces additional memory overhead due to concurrent witness generation and proof computation. 
%To address this, we introduce an automated circuit partitioning algorithm that divides large circuits into smaller subcircuits, effectively reducing memory consumption while minimizing resource contention. 
%Our approach tackles both computational and memory efficiency challenges, providing a more adaptable solution for ZK-SNARK systems. 
Experimental results confirmed that our framework effectively improves the resource utilization and proof generation speed.
\end{abstract}

\maketitle % should come after the abstract
\pagestyle{plain} % should come right after \maketitle

\section{Introduction}
Zero-knowledge proof, a prevailing cryptographic technique, allows one party to convince another of the correctness of a statement while not leaking any valuable secret of it. \zk (Zero-Knowledge Succinct Non-Interactive Argument of Knowledge), the most popular protocol, is widely used in industry to protect privacy and secure data, as it additionally avoids the interaction between prover and verifier and generates proof with constant size. For instance, \zk  
is used in outsourced computing platforms to force the service providers to accomplish the target task delegated by users as expected. The succinct feature of \zk's proof enables fast verification on the user side. To pursue better service, the provider's mission is to respond to all requests by generating corresponding proofs as quickly as possible. However, the inherent performance bottleneck in proof generation significantly stops the throughput improvement.

Typically, the proof generation for \zk consists of two phases:  \textit{witness generation(WG)} and \textit{proof computation(PC)}. The WG  constructs the necessary inputs, known as the witness, satisfying a specified predication and the PC produces a succinct cryptographic proof based on the witness and some other data. Recently, the optimization for the PC phase has made a great achievement, successfully accelerating the execution through various parallel solutions \cite{botrel2023faster, chen2024load, zhang2021pipezk, ma2023gzkp, ji2024accelerating}. Conversely, the development of PC optimization is deeply under our expectation, because the WG phase is closely associated with the specific logic of the task to be proved, which prevents general optimization in advance. 
Consequently, the WG naturally becomes a new performance bottleneck for proof generation.

%Optimization of PCs, whether through the underlying implementation of algorithms \cite{botrel2023faster, chen2024load} or based on hardware acceleration \cite{zhang2021pipezk, ma2023gzkp, ji2024accelerating}, is relatively easy as they are both general-purpose optimisations that are not related to specific circuit logic.

%However, optimising the PC phase alone is not sufficient for a process that incorporates both WG and PC for proof generation. In the process of proof generation, after the PC is optimised fast enough, the WG naturally becomes a new performance bottleneck for the system. At this point, even if more physical resources are allocated, the system performance cannot be further improved. And the WG is closely related to the specific circuit logic, it is difficult to do general optimisation in advance.

\begin{figure}[t]
    \centering
    \includegraphics[width=0.45\textwidth]{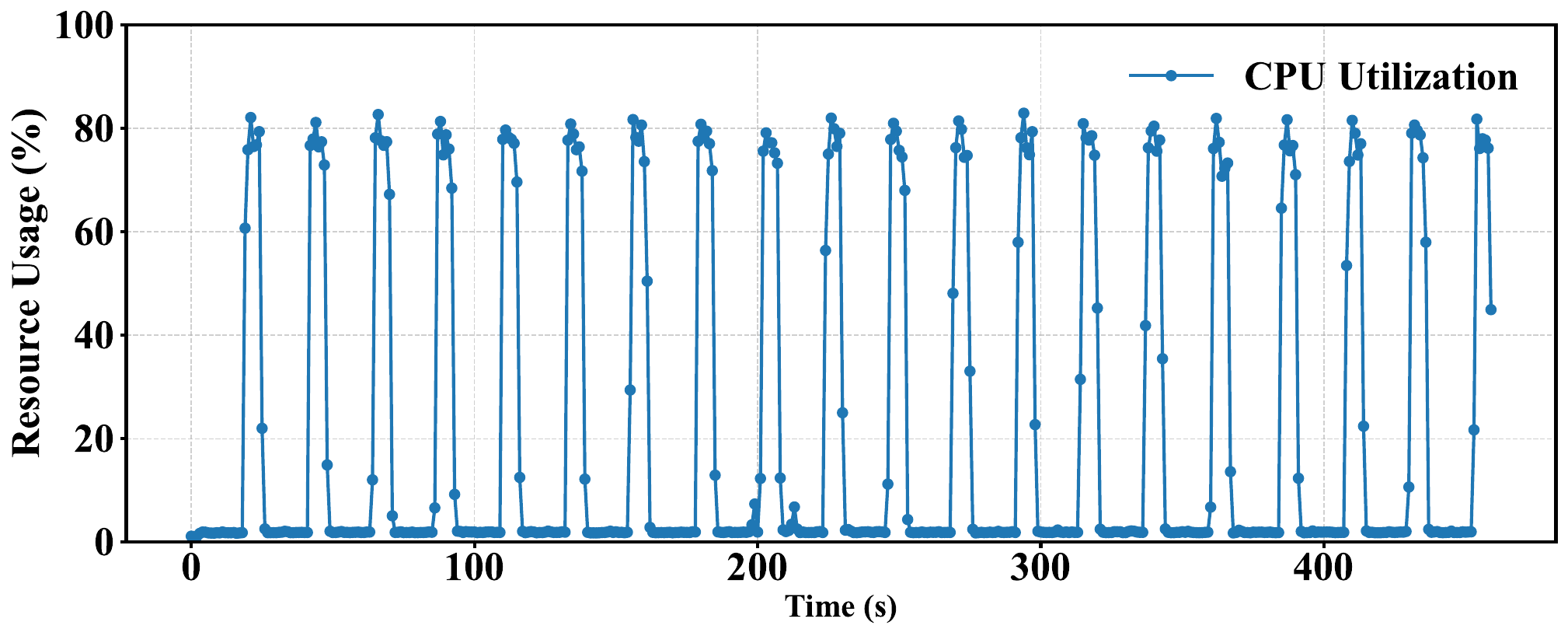}
    \caption{CPU usage with continuous tasks over time.}
    \label{fig:CPU_track}
\end{figure}

\begin{figure}[t]
    \centering
    \begin{subfigure}[b]{0.22\textwidth}
        \centering
        \includegraphics[width=\textwidth]{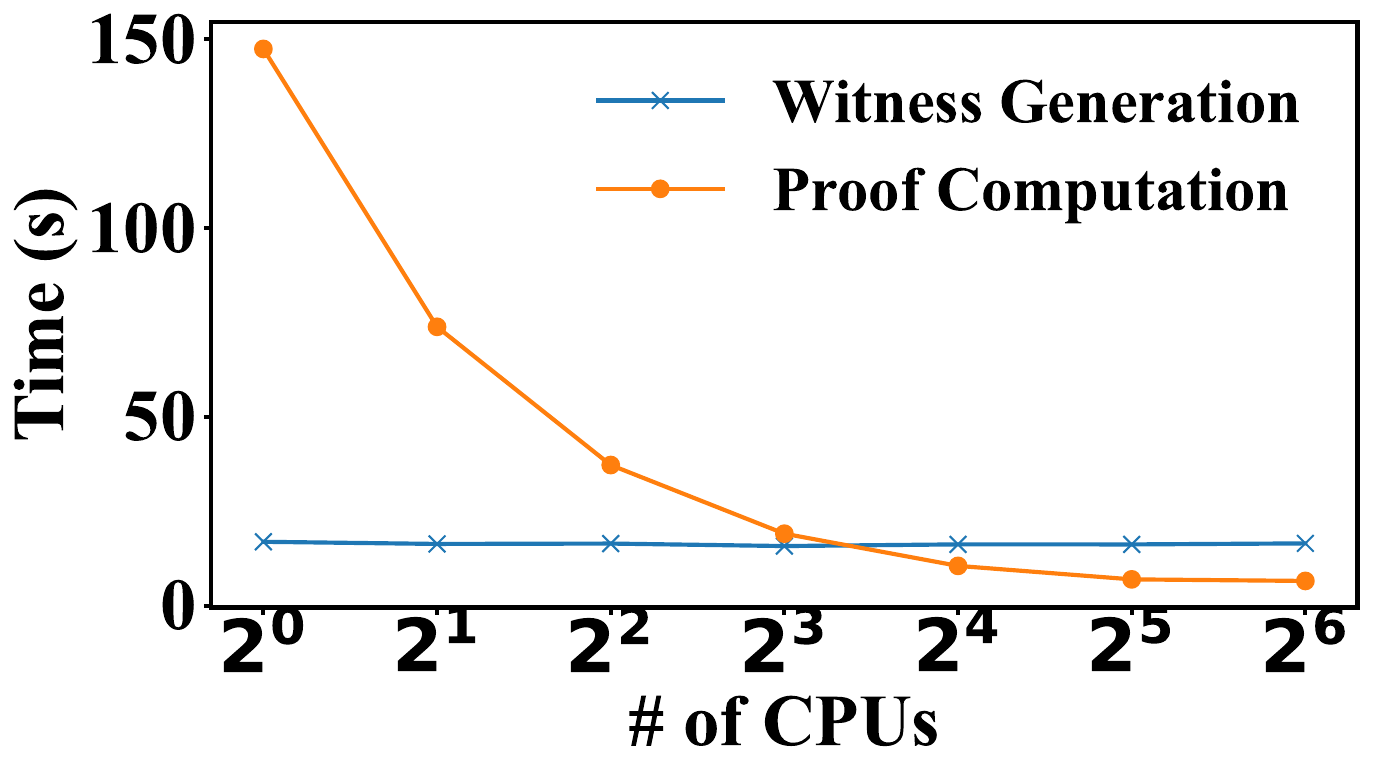}
        \captionsetup{justification=centering}  % Center the caption
        \caption{Time Cost}
        \label{fig:solve_vs_prove_a}
    \end{subfigure}
    % \hfill
        \hspace{0.01\textwidth}
    \begin{subfigure}[b]{0.22\textwidth}
        \centering
        \includegraphics[width=\textwidth]{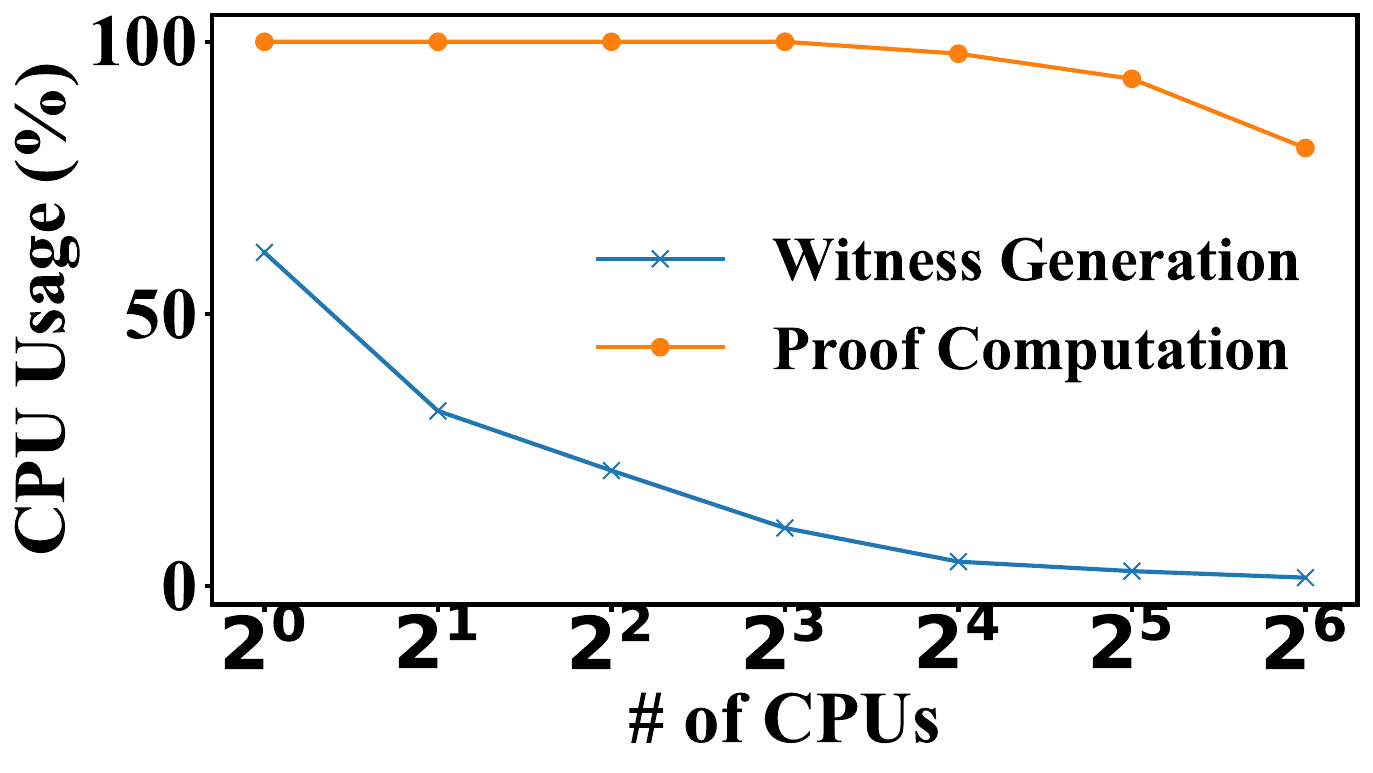}
        \captionsetup{justification=centering}  % Center the caption
        \caption{CPU Utilization}
        \label{fig:solve_vs_prove_b}
    \end{subfigure}
    \caption{Performance for witness generation and proof computation across different numbers of CPUs. }
    \label{fig:solve_vs_prove}
\end{figure}

To make a better understanding of this issue, we tested a real example of the \zk, where the logic is expressed by a \emph{circuit}. Basically, we monitored the CPU usage over time against a stream of continuous tasks of proof generation. 
These tasks all rely on identical circuits—a linear recurrence circuit with 60 million constraints (refer to \cref{sec:evaluation} for details). As shown in \cref{fig:CPU_track}, the CPU utilization 
exhibits intense fluctuations over time
and commonly remains at an extremely low level. Despite there are sufficient tasks to run,  CPU utilization still exhibits a consistent cyclical pattern, indicating inefficiencies in resource utilization. 
In light of this, we conducted further tests focusing on the two phases separately.  \cref{fig:solve_vs_prove} showcases that increasing CPU resources significantly accelerates proof computation, while the latency of witness generation remains largely unaffected regardless of the CPU number. This inefficiency highlights that witness generation has become the bottleneck in the {\zk} system, preventing the optimal utilization of available computational resources.

% \begin{figure}[t]
% \centering
% \includegraphics[width=0.48\textwidth]{asplos25-templates/figures/solve_vs_prove.pdf}
% \caption{Performance for witness generation and proof computation across different numbers of CPUs.}
% \label{fig:solve_vs_prove}
% \end{figure}
% cpu利用率要上去不需要partition，partition对内存的减少需要无pk（有pk就是单纯减少prove memory），就是一次性，cpu那边需要堆pk，然后partition针对一个电路，目前后续pipeline针对证明流
%%%之前讨论的分别对CPU和内存的要求不一样还存在吗？

%Fortunately, ...

\paragraph{Our solution} In this paper, we propose \system, a scalable framework for optimal resource utilization in \zk systems.
To increase the CPU utilization rate, \system leverages the \emph{pipeline} to make the proof generation for different tasks interleave. To this end, \system decouples the witness generation phase and proof computation phase and assigns them for adjacent tasks to distinct computing units. Ideally, all computing units should be occupied by tasks simultaneously, maximizing CPU utilization. However, this pipeline design comes up with two severe issues to be solved: \emph{increased memory consumption} and \emph{imbalanced execution costs}.  

\paragraph{Increased memory consumption} As more tasks, especially the witness generation, run concurrently, memory consumption grows at an incredible speed—which becomes not affordable to systems. To mitigate this, \system adopts a topological sort-based greedy partitioning algorithm that breaks down large circuits into smaller subcircuits, which can be executed sequentially and separately.  \system guarantees that by executing these subcircuits in a certain order, the final result is equivalent to that of directly running the full circuit. This makes executing large-scale circuits with limited memory resources possible in practice.

%reducing the overall memory required to complete the entire computation, while ensuring that the total proof generation time remains nearly unchanged.

%By combining partitioning with the pipeline model, we enable faster proof generation while keeping memory usage within acceptable limits, offering a scalable and efficient solution for verifiable computation involving large-scale zero-knowledge proof circuits.

\paragraph{Imbalanced execution costs}
The benefit of pipeline is maximized when the costs of witness generation and proof computation for each subcircuit are approximately equal. However, the performance of witness generation is influenced by the circuit and resource configuration and is commonly not comparable with proof computation, resulting in a skewed workload for both phases. Therefore, we propose a scalable framework in \system, where the ability of each phase is scalable and configurable. By tuning the resources assigned to both phases, \system can achieve a balanced workload across phases. 

%the cost of witness generation to proof computation varies depending on the circuit and resource configuration, meaning that the benefits of a simple pipeline model may be limited if the concurrency is not configurable. As a result, we propose a configurable pipeline model that allows multiple witness generation tasks to be executed concurrently with a single proof computation task.

%Although WGs and PCs cannot be brought up in parallel for the same circuit, for a batch of circuits, full consideration can be given to further enhancing the resource utilisation of the system and improving the throughput capability of the system by pipelining the WG and PC phases between multiple circuits.
%A natural solution is to use \textbf{pipeline} techniques to address this challenge. Since proof computation and witness generation are sequential but independent processes, they inherently fit into a pipeline model.  However, the cost of witness generation to proof computation varies depending on the circuit and resource configuration, meaning that the benefits of a simple pipeline model may be limited if the concurrency is not configurable. As a result, we propose a configurable pipeline model that allows multiple witness generation tasks to be executed concurrently with a single proof computation task.

%However, the pipeline model introduces a new issue: \textbf{Increased Memory Overhead}. 

In summary, our contribution to this paper includes:
\begin{itemize}[leftmargin=*]
    \item[1)]  We propose the pipeline technique to increase the CPU utilization in \system tasks. Moreover, we introduce an automatic circuit partitioning algorithm that divides large circuits into sequential subcircuits, which decreases the memory usage to execute large-scale circuits. 
    \item[2)]    We decouple the witness generation phase and proof computation phase for each subcircuit and enable them scalable separately. This model balances the costs of both phases,  optimizing parallelism granularity for faster proof generation.
    \item[3)] We implement \emph{\system}, which integrates the proposed technologies and conducts extensive experimental evaluations. The results demonstrate that \system significantly improves resource utilization and proof generation speed under controllable memory usage.
\end{itemize}
%\raggedbottom

The remainder of the paper is structured as follows. \cref{sec:background} covers the background on \zk and the motivation for this work. \cref{sec:overview} overviews {\system}'s design, and \crefrange{sec:partitioner}{sec:scheduler} detail its specific components.  \cref{sec:evaluation} evaluates \system's performance  while \cref{sec:related_work} reviews related literature. Finally, \cref{sec:conclusion} concludes the paper.
\section{Background and Motivation} \label{sec:background}

This section introduces some necessary background about \zk and analyzes the issues that motivate this work.

\subsection{{\zk}} \label{subsec:zk}
% todo 这里的参考文献

\begin{figure}[t]
    \centering
    \includegraphics[width=0.48\textwidth]{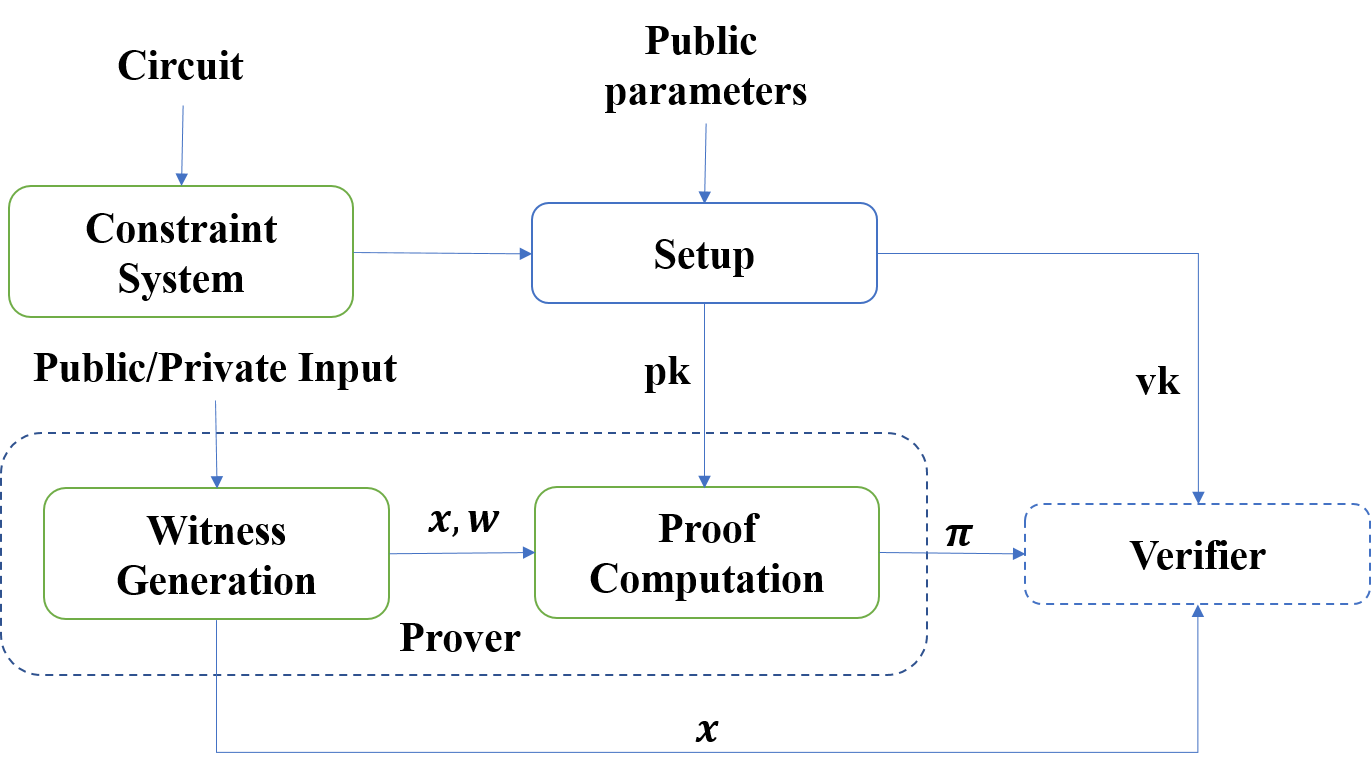}
    \caption{The complete process of a \zk protocol.}
    \label{fig:zk-snarks}
    \vspace{10pt}
\end{figure}

Zero-knowledge proofs enable one party, the prover, to demonstrate the validity of a statement to another party, the verifier, without revealing any information about the statement.
\zk \cite{bitansky2013succinct, chiesa2020marlin} is one of the well-known Zero-Knowledge protocols, which has been widely adopted in various real-world applications. For example, the blockchains, especially cryptocurrency systems, employ it to validate transactions while keeping details confidential for privacy protection \cite{sasson2014zerocash,danezis2013pinocchio,filecoin2022,quorum2022,guo2024zkcross,kosba2016hawk,mina2022,qedit2017}.
Furthermore, {\zk} can enable off-chain execution and on-chain verification based on the succinct proofs, mitigating the on-chain transaction pressure \cite{xie2022zkbridge,mina2022,xu2024exploring}.
Beyond this, \zk is applied in verifiable outsource system\cite{wahby2014efficient,costello2015geppetto,delignat2016cinderella}, such as database outsourcing \cite{li2023zksql,zhang2017zero} and privacy-preserving machine learning \cite{liu2021zkcnn,chen2024zkml}, promising result correctness and data privacy.

\begin{figure}[t]
    \centering
    \includegraphics[width=0.44\textwidth]{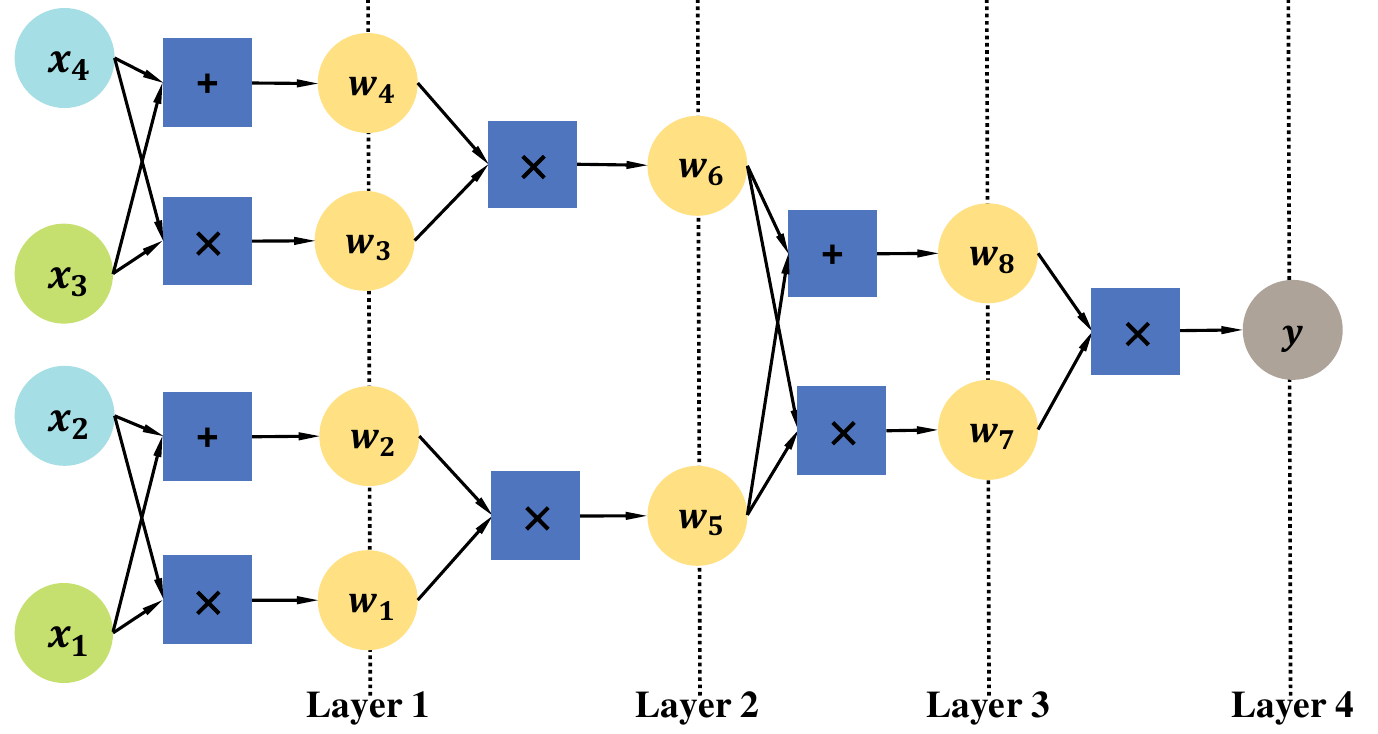}
    \caption{An example of the \zk circuit. }
    \label{fig:ex_zk_circuit}
\end{figure}

In {\zk}, the computation logic is expressed by a \emph{circuit} as examplified in \cref{fig:ex_zk_circuit}, where each step of the computation is converted to a logical operation. To guarantee the circuit execution meets our expectation, some \emph{constraints}, i.e., a set of equations/inequations, are imposed on the inputs and outputs.  \Cref{fig:zk-snarks} shows a complete process of a \zk protocol:
%, which can be formally expressed as a three-tuple of polynomial time algorithms
\begin{itemize}[leftmargin=*]
    \item $Setup(1^\lambda,F)\to(pk_F,vk_F)$. This phase takes a predicate $F$ (i.e., ) as inputs and generates a proving key $pk_F$ and a verification key $vk_F$. The setup is performed once per circuit and can be reused for multiple proofs about $F$.
        
    \item $Prove(x,w,pk_F)\to\pi$. This phase generates a proof $\pi$ using the instance $x$, witness $w$, and proving key $pk_F$. The proof $\pi$ attests that $w$ satisfies $F$ with respect to $x$, without revealing any information about $w$.
    
    \item $Verify(x,\pi,vk_F)\to\{0,1\}$. This phase verifies whether proof $\pi$ is valid, for instance, $x$, using the verification key $vk_F$. It can efficiently check the proof without the need to recomput $F$.
\end{itemize}

Generally, the setup phase is a one-time operation, and the verify phase is performed on the verifier's side. Thus, the performance for them is not our concern. The prove phase, the focus of this work, commonly consumes tremendous computational resources. Particularly, the prove phase in \zk consists of two major components: \emph{witness generation} and \emph{proof computation}.

\paragraph{Witness generation} Witness generation builds the necessary inputs required for proof computation, known as the witness $w$ satisfying the predicate $F$, against a given instance. 
%It prepares the foundational data required for the proof without revealing sensitive information. 
The intermediate values are computed during witness generation to satisfy the circuit's constraints. The circuit can be regarded as a layered graph, and the computation is processed layer-by-layer as depicted in \cref{fig:ex_zk_circuit}. Specifically, the nodes in each layer depend on the outputs from the previous layer. Within each layer, all nodes are completely independent, implying the parallel computation for nodes in the same layer. Thus, the parallelism potential of the witness generation phase is highly related to the circuit's structure.

% 这里NTT还是要说...因为现在NTT用的比FFT多,FFT算是ZKP刚提出来的时候的标配，NTT比它更新一点，现在更常用
\paragraph{Proof computation} Proof computation produces a succinct cryptographic proof from the witness based on the proving key and the circuit constraints. This phase involves several computationally intensive operations, including the polynomial (POLY) evaluation and the multi-scalar multiplication (MSM). During the POLY stage, the Fast Fourier Transform (FFT) or its variant, the Number Theoretic Transform (NTT), is applied to efficiently evaluate polynomials over a large number of points. The subsequent MSM stage combines multiple elliptic curve points with their corresponding scalar values. The MSM stage is particularly resource-intensive, often accounting for up to 70\% of the total proof generation time in CPU-based \zk implementations \cite{ma2023gzkp}.

%It is important to note that while proof computation depends on the completion of its corresponding witness generation, the witness generation and proof computation of separate tasks are entirely independent of one another.

\begin{figure*}[t]
%\vspace{10pt}
\centering
\includegraphics[width=\textwidth]{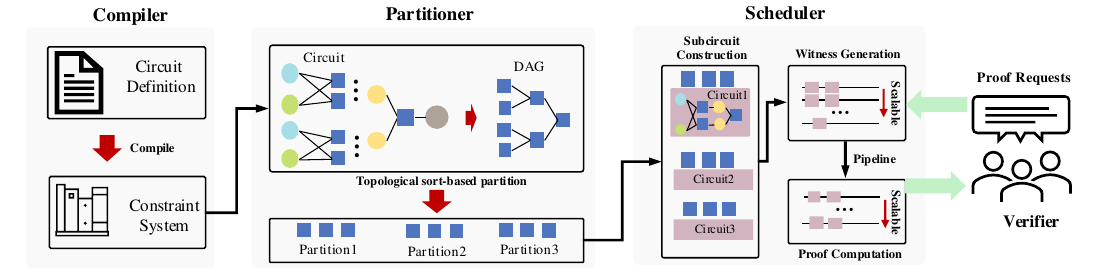}
\caption{System architecture and overall workflow of \system.}
\label{fig:overview}
\end{figure*}

\vspace{-10pt}

\subsection{Motivation}

Although \zk has made significant progress in various aspects, two challenges still hinder its applications for large-scale circuits.

\paragraph{Inefficiencies in witness generation} 
Significant efforts, e.g., Gnark \cite{gnark}, have been dedicated to optimizing the proof computation phase, enhancing resource utilization, and accelerating proof generation. However, the witness generation phase that draws less attention has gradually become a bottleneck in the overall system efficiency. This is because the parallelism of witness generation directly inherits from the circuit structure, which is difficult to accelerate. As a result, the average resource utilization remains low even with highly optimized proof computation. This situation will escalate in scenarios where proof generation is continuously invoked, such as verifiable computation platforms handling concurrent user requests. Therefore, how to increase resource utilization becomes a critical issue for such platforms.

%This imbalance between witness generation and proof computation leads to suboptimal CPU usage, which is detrimental in high-throughput environments. Particularly, in scenarios where a continuous stream of proofs needs to be generated, such as verifiable computation platforms handling concurrent user requests, the inefficiencies in witness generation become a critical issue.

% 按照上一次的讨论，在motivation里是把内存单独拿出来的，然后下面的逻辑就是1. zkp本身要内存很多；2. 这种内存大的问题会顺带影响到一些并行的优化，就是说本来跑一个就要很大内存了，跑多个更多了；这样能挂上pipeline，然后目前的intro里的逻辑是先说利用率->pipe,然后pipe->内存太大->partition
% 一种可能是把内存也作为一个单独的贡献，就是说，我们通过partition解决了内存问题，逻辑上来说没问题，实际情况是各个子电路的pk是要同时存在内存的，因为不可能只用一次都不用了，（尝试过持久化，读进来也很慢很慢，分钟级别），pk太大了，所有子电路的pk大小总和可以认为和原来的一个pk差不多，所以这部分内存没动，减少的是prove过程的内存，这个确实1/n，但这部分内存比pk小不少，大概是40G：20G，然后partition是动了那20G；可以看实验，实验是区分了total memory(包括pk)和Prove memory（不包括pk的），然后partition的实验里total memory不算堆叠pk，pipeline实验里算了堆叠pk；

\paragraph{Memory overhead} 
\zk circuits frequently comprise millions of constraints, leading to substantial memory requirements. This large memory footprint constrains the scalability of \zk systems and undermines the potential of parallel optimizations in proof generation. To address this, SPLIT \cite{qi2023split} introduced an approach that partitions circuits into smaller subcircuits and executes them sequentially. However, this method relies on manual circuit division, impractical in current in-production \zk circuits. As the size and complexity of the circuits increase, manual division is even prone to errors. Consequently, there is a pressing need for a general and automatic circuit partitioning algorithm specifically designed for single-machine environments.

\section{Overview}\label{sec:overview}
In this section, we provide an overview of \system, covering its system architecture as well as the foundational model on which it operates. 

\subsection{System Model}

\system operates based on the following three fundamental assumptions.

\paragraph{Continuous requests} The advancement of \system relies on maximizing the overall throughput for proof generation of \zk towards a stream of requests rather than optimizing the performance of a single run. Thus, \system aims to increase the peak throughput when dealing with continuous requests of proof generation from users, e.g., in the outsourcing platform. 

\paragraph{CPU-based optimization} \system focuses on optimizing the performance of a CPU-based  {\zk} implementation. The works that leverage GPUs to enhance \zk systems \cite{chen2017big,dai2016cuhe,goey2021accelerating,kim2020accelerating},  are orthogonal to our work and \system still can adapt to these works after applying some necessary modifications. 

\paragraph{Public variables} The intermediate computed results of circuits
can be safely exposed as public variables. It means that the execution outcome itself may be revealed to the public.  \system suits for scenarios where the primary concern is to protect the private inputs. 
For example, in most ZK-Rollup implementations, the focus is on leveraging the succinctness of SNARKs to verify off-chain computations rather than on ensuring zero-knowledge property.

%Even in scenarios where zero-knowledge properties are required, such as when protecting user privacy, the primary concern is to keep the private inputs (e.g., user transaction details) confidential. The intermediate computation results are often not sensitive and can be safely exposed without compromising privacy.

\subsection{System Architecture}

\Cref{fig:overview} depicts the system architecture of \system, encompassing three core components: \emph{Compiler}, \emph{Partitioner}, and \emph{Scheduler}, which collectively facilitate the generation of \zk proofs.

%These components collectively facilitate the generation of ZK-SNARK proofs by breaking down the process into distinct, manageable tasks.

\paragraph{Compiler}  
The compiler is responsible for compiling the ZKP circuits, converted from the original programs written in high-level languages, into constraint systems, such as R1CS \cite{parno2016pinocchio}, Plonkish \cite{gabizon2019plonk} and AIR \cite{ben2019scalable}. \system is irrespective of the concrete implementation of the compiler. A variety of existing tools, e.g., Circom \cite{circom}, Gnark \cite{gnark}, and Libsnark \cite{libsnark},  can play the role of the compiler in \system. 

\paragraph{Partitioner}  
The partitioner divides each large \zk circuit into multiple subcircuits. Executing these smaller sub-circuits sequentially allows resource-limited machines to execute the large-scale circuit. Moreover, each subcircuit is a basic execution unit scheduled by the scheduler. \system, inherently,  models a circuit as a Directed Acyclic Graph (DAG) and transforms circuit division into a DAG partition problem with two goals.  First, the partition must be serializable, meaning the results of executing the subcircuits in a topological order are equivalent to the original full circuit. Second, the load of each circuit, in terms of the number of nodes and across-partition dependencies, should be approximately equal, ensuring balanced workloads. We refer to this process as a \emph{balanced serializable circuit partition} in this work, completed via a greedy algorithm based on topological sorting.

\paragraph{Scheduler}  
The scheduler coordinates the execution of the partitioned subcircuits, ensuring they follow the correct order in accordance with the circuit’s dependencies. With continuous proof generation requests submitted to \system, the scheduler decouples the witness generation and proof computation for each subcircuit from different requests and applies the pipeline design to process execution across them. Ideally, this will increase the parallelism between processes of different requests.
However, the costs for witness generation and proof computation are not balanced, offsetting the benefits brought by the pipeline. To address this issue,  \system enables the capability of both configurable by a scalable framework. In this scenario, we can dynamically assign computational resources to both phases, ensuring their latencies are close to maximize the pipeline's potential.

We will elaborate on the designs of the partitioner and scheduler in the following two sections.

\section{Partitioner} \label{sec:partitioner}

In this section, we present the partitioner in detail, which divides each large \zk circuit into smaller sub-circuits, which can be executed serially and independently while promising the correctness of results. 

%The primary goal is to provide flexible control over memory usage during the proof generation  phase, while maintaining the correctness of the circuit logic and ensuring that the overall execution time remains nearly the same.

\subsection{Problem Formalization}
% We start by formally defining the concept of \textit{Serializable Circuit Partitioning}.

A \zk circuit $\mathcal{F}$ can be defined as,
$\mathcal{F}(I_{p}, I_{s}) = (O_{p}, O_{s})$. Here, $I_{p}$ and $I_{s}$ represent the public and secret inputs, respectively. Similarly, $O_{p}$ and  $O_{s}$ are the public and secret outputs. The secret inputs $I_{s}$ are used internally by the prover, while the secret outputs $O_{s}$ should remain hidden and are not revealed to the verifier.
Apart from the inputs/outputs, another important component in a circuit is \emph{constraints}.  Constraints are sets of equations or inequalities that verify the correctness of the circuit.  They describe the relationship between inputs and outputs. In \cref{fig:ex_zk_circuit},  the nodes  $x_1 \sim x_4$ are inputs, and $y$ is the public output. Each arithmetic constraint corresponds to an output, where private output like $w_1 \sim w_8$ is hidden as part of the intermediate computation, while the final public output $y$ is revealed at the end of the process.

The goal of the partitioner is to divide the original circuit $\mathcal{F}$ into subcircuits  $\{F_1, \ldots, F_k\}$, such that each sub-circuit can be executed sequentially. To this end, \system extracts a \textit{Constraint Dependency Graph (CDG)} from the circuit and performs the partitioning onto it. 

% 原本带有Input的定义
% \begin{definition}[Constraint Dependency Graph]\label{def:cdg}

% A \textit{Constraint Dependency Graph (CDG)} is a directed acyclic graph  $G = (V, E)$. Here, each vertex in $V$ represents a constraint or input of the original circuit.  For each edge $(u, v) \in E$, the output of constraint $u$ is forwarded to the input of $v$. 
% \end{definition}
\begin{definition}[Constraint Dependency Graph]\label{def:cdg}
A \textit{Constraint Dependency Graph (CDG)} is a directed acyclic graph  $G = (V, E)$. Each vertex in $V$ represents a constraint of the original circuit.  For each edge $(u, v) \in E$, the output of constraint $u$ is forwarded to the input of $v$. 
\end{definition}

\begin{figure}[t]
    \centering
    \includegraphics[width=0.48\textwidth]{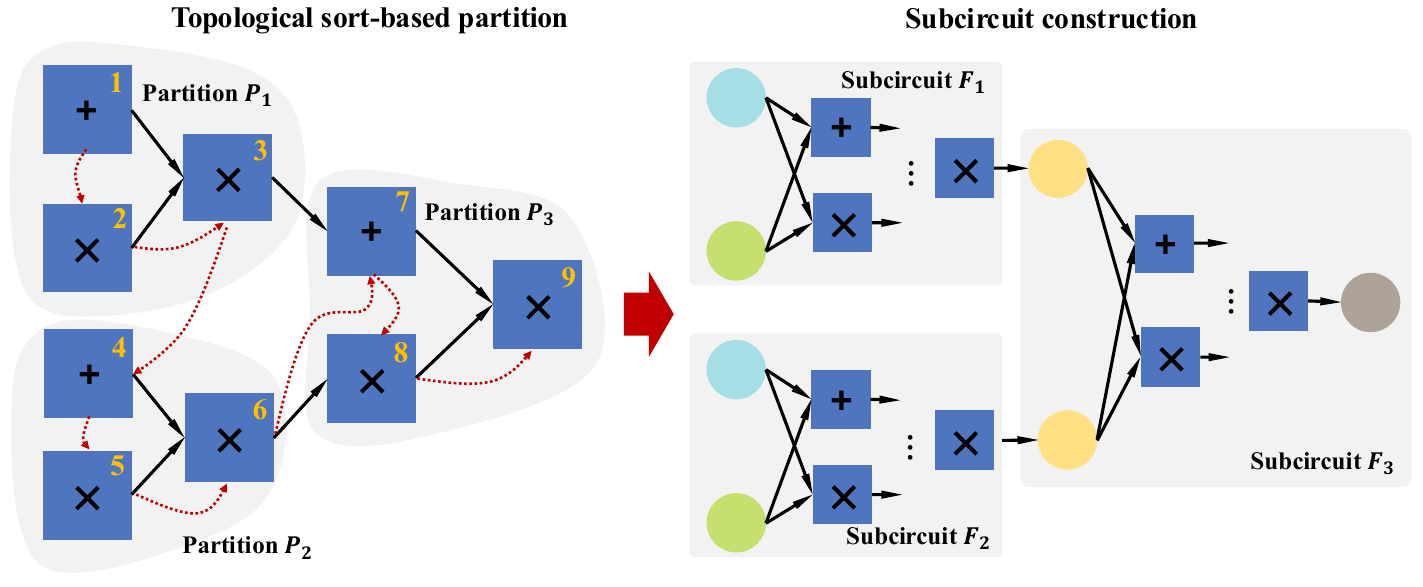}
    \caption{An example of serializable circuit partitioning and the subcircuit construction. }
    \label{fig:circuit_partition}
\end{figure}

A CDG captures the dependencies between all constraints. The left-hand side of \cref{fig:circuit_partition} depicts the CDG for the circuit in \ref{fig:ex_zk_circuit}. Based on CDG, the circuit partitioning problem is transformed into a graph partitioning problem. For a CDG $G$, a partition  is $\mathcal{P}=\{V_1, V_2, \ldots, V_k\}$, such that $G.V=\cup_{1\le i \le k}V_i$ and $V_i \cap V_j = \emptyset, \forall i\ne j$. The nodes in $V_i$ correspond to the constraints in each subcircuit $F_i$, and the edges between nodes in $V_i$ remain the connection relationships between constraints\footnote{In this section if there is no ambiguity, we use a subcircuit and partition interchangeably. }.
Besides, we require the partition to satisfy two conditions: \emph{serializable}, which promises the sequential and independent execution of each subcircuit; \emph{balanced}, which ensures the load of each subcircuit,  in terms of computational resources to pay,  is bounded and approximately equal.

\paragraph{Serializable partition} If a partition $\mathcal{P}$ is serializable, it means each $V_i$ only depends on the set $V_j$ proceeding itself, i.e., $j<i$. Specifically, $V_i$ depends on $V_j$ when there exists edge $(v, u) \in G.E$, such that $u \in V_i$ and $v \in V_j$. Given a serializable partition, \system can execute all sub-circuits (or partition) sequentially because when $V_i$ is going to be executed, all the sub-circuits depended on have been finished.

\paragraph{Balanced partition} The load $L(V_i)$ of a sub-circuit $V_i$ is defined as: 
\begin{equation}\label{equ:circut_load}
L(V_i) = |V_i| + \sum_{j < i} \left| \{  (u, v) \in E \mid u \in V_j, v \in V_i\} \right|     
\end{equation}

Commonly, the load of each partition is proportional to the resources used to execute it. 
Therefore, a balanced partition aims to minimize the maximum load across partitions: 
$\min \max_{1\leq i \leq k}(L(V_i))$. 

%balanced里面，减少shared variables的一个目的是内存，另一个更主要的目的是减少验证代价，因为shared variables是否在两个子电路上匹配需要验证方额外验证
In \cref{equ:circut_load}, two parts contribute to the subcircuit load $L(V_i)$: \romannumeral1) number of nodes in $V_i$ and \romannumeral2) number of cross-partition edges between $V_i$ and other partitions. The number of nodes in a sub-circuit directly impacts memory usage during proof generation. On the other hand, more cross-partition edges indicate more \emph{shared variables}, intermediate results, should be shared between constraints, increasing memory usage during the witness generation phase. Moreover, these shared variables also need to be verified, consuming additional computation. 
Therefore, we also take it into account of the load. When dealing with large-scale circuits, we should limit the sub-circuit size and shared variable number, making the memory consumption for subcircuits can fit the real equipped resources.

\subsection{Topological Sort-base Greedy Partition}

The circuit partition problem defined yet is a highly complex problem. To address this problem, we propose a greedy partition algorithm based on the topological sort.  We guarantee the serializability of partition by topological sorting and achieve the balance via two greedy strategies. 

Topological sorting performs a linear ordering on a DAG, which motivates us to design a serializable partition. Given a topological sorting $ v_1, v_2, \dots, v_N$ for the graph $G$, the serializable partition for $V_i$ is defined as: 
$$V_i = \left\{ v_t \mid (i-1) \cdot \frac{N}{k} < t \leq i \cdot \frac{N}{k} \right\}$$
Apparently, a node $v_i$ must proceed $v_j$ in the ordering if an edge $(v_i, v_j) \in G.E$ exists. Then, nodes in $V_i$ only depend on the nodes in proceeding partitions $V_j$ ($j\le i$) according to this strategy, equivalently promising the serializability of $\mathcal{P}$.

The above approach only solves one requirement of partitioning, and the remaining focus is on achieving balance. However,  the challenge lies in finding a topological sorting that minimizes the number of shared variables across partitions. Instead, we propose a greedy algorithm to solve this problem. Before delving into the details, we have to introduce two concepts first.

\begin{definition}[Depth]
The depth $d(v)$ of a node $v \in G.V$ is the length of the longest path from any node without in-edges to $v$.
\end{definition}

\begin{definition}[Out-Degree]
The out-degree $g(v)$ of a node $v \in V$ is the number of edges directed outward from  $v$, indicating how many other constraints depend on $v$'s output.
\end{definition}
In \cref{fig:circuit_partition}, the depth and out-degree of node numbered with 7 are 3 and 1, respectively. During the topological sorting, when visiting a node, we follow two heuristic strategies to explore its children nodes:
\begin{itemize}[leftmargin=*]
    \item Nodes with smaller depth are prioritized as they depend on fewer constraints.  
    This allows shallower nodes to be processed earlier and the deeper dependencies are grouped together in later partitions.

    \item Nodes with smaller out-degrees are prioritized because they affect fewer other constraints. By this approach, the impact of the current partition on future partitions is minimized, reducing shared variables between them.
\end{itemize}

\system integrates a DFS-based topological sorting to order nodes in original $G$. In particular, DFS ensures that a node is processed only after its dependencies have been fully resolved. 
%It more effectively captures the dependencies between nodes, reducing the number of shared public variables between partitions.
When traversing from a node,  we explore its neighbored nodes with priorities according to the aforementioned two strategies to achieve a balanced partitioning.

In \cref{fig:circuit_partition}, the numbers attached to nodes indicate their order generated by the topological sorting with proposed greedy strategies. Then, the CDG is partitioned into three parts. Actually, we still have to recover the subcircuits with respect to partition. The scheduler does this, which is detailed in the next section.   

\begin{algorithm}[t]
\caption{Topological Sort-based Greedy Partition }
\label{alg:partition}

\require{\emph{Set of constraints $C$, partition number $k$}}

\ensure{\emph{Serializable partition $\mathcal{P} = \{V_1, V_2, \dots, V_k\}$}}\\

$G^{-1}\gets {\rm Rev\_CDG}(C)$ \tcp{Construct Rev-CDG} \label{stmt:rev_cdg_construct}
$L_{root}\gets$ nodes in $G^{-1}$ without incoming edges \label{stmt:root}\\
Initialize partition set $\mathcal{P} \gets \emptyset$, $V_p\gets \emptyset$\\
Initialize $s \gets \left\lceil {|V|}/{k} \right\rceil$ \label{stmt:partition_size}\\

\For{{\rm each root node $r \in L_{root}$} \label{stmt:root_partition_start}} {
    {Rec\_Partition}{$(r, \mathcal{P}, V_p, G^{-1}, s)$} \label{stmt:root_partition_end}
}
    \Return{$\mathcal{P}$} \label{stmt:return_partition}\\
\vspace{5pt}
\function{\emph{Rev\_CDG($C$)}} {
    Initialize a Rev-CDG $G^{-1} \gets (V=\emptyset, E^{-1}=\emptyset)$ \label{stmt:rev_cdg_init}\\
    \For{\rm each constraint $c \in C$}{
        $V\gets V\cup \{v_c\}$ \label{stmt:add_constraint}
    }
    \For{\rm each pair of constraints  $(c_1, c_2) \in C$ \label{stmt:add_edge_start}} {
        \If{\rm output of \( c_1 \) is an input to \( c_2 \)} {
            $E^{-1}\gets E^{-1}\cup \{(v_{c_2}, v_{c_1})\}$\\
            \tcc{Update depth and out-degree}
            $d(v_{c_2}) \gets max\left\{d(v_{c_1}) + 1, d(v_{c_2})\right\}$\\
            $g(v_{c_1})\gets g(v_{c_1})+1$ \label{stmt:add_edge_end}
        }
        
    }
    \Return{$G^{-1}$}
}

\vspace{5pt}
\function{\emph{Rec\_Partition($v,\mathcal{P}, V_p  G^{-1}, s$)} \label{stmt:rec_partition_start}} {    
    \If{ {\rm $v$ has been visited}} {
        \Return
    }
     Mark $v$ as visited\\
     \tcc{Visit nodes according to their priorities}
    \For{ {\rm $u \in  Sorted\_Children(v, G^{-1})$} \label{stmt:explore_children_start}} {
        {Rec\_Partition}{($u, \mathcal{P}, V_p, G^{-1}, s$)} \label{stmt:explore_children_end}
    }
    $V_p\gets V_p\cup \{v\}$ \tcp{Add to current partition}
    \If{$|V_p| = s$ \label{stmt:add_new_partition_start}} {
        $\mathcal{P}\gets \mathcal{P}\cup \{V_p\}$ \tcp{Add new partition} \label{stmt:add_to_partition}
        $V_p\gets \emptyset$ \label{stmt:add_new_partition_end}\label{stmt:rec_partition_end}
    }
}

\end{algorithm}

\subsection{Algorithm}
% 逻辑： 要找到一个拓扑排序-> DFS-based得到拓扑排序-> DFS(后序遍历)对等位节点的遍历顺序和结果直接绑定-> 基于depth和out-degree选择先遍历谁   
In this subsection, we present the details of our partition algorithm. Our algorithm performs the topological sorting against a \emph{Reversed Constraint Dependency Graph}, defined in \cref{def:rev_CDG}, instead of the original CDG. 
\begin{definition}[Rev-CDG] \label{def:rev_CDG}
Given a CDG \( G = (V, E) \), its corresponding Reversed Constraint Dependency Graph \( G^{-1} = (V, E^{-1}) \) satisfied $\forall (u, v) \in E$, \( (v, u) \in E^{-1} \), and \( |E| = |E^{-1}| \). 
\label{def:rcdg}
\end{definition}
\noindent This choice stems from the need to ensure that a node $v$ should be assigned to a partition until the assignment is done for all nodes it depends on, which are the source nodes for incoming edges to $v$. However, only outgoing edges are available in the original CDG. By reversing the graph $G$, these nodes become the children of $v$ in the $G^{-1}$. This reversal enables us to conduct a post-order traversal to order nodes topologically, where a node is added to the list once all its children have been visited.

% 下面就是RCDG的构建过程和Partition算法
\Cref{alg:partition} presents the details of the topological sort-based greedy partition algorithm. It takes a set of constraints $C$ derived from the circuit directly and the expected number $k$ of partitions as inputs. First, it constructs the reversed CDG $G^{-1}$ by invoking function Rev\_CDG$()$ (Line \ref{stmt:rev_cdg_construct}). The Rev\_CDG$()$ initializes an empty Rev-CDG $G^{-1}$ and each constraint in $C$ correspond a node in $G^{-1}$ (Lines \ref{stmt:rev_cdg_init}-\ref{stmt:add_constraint}). Then, each dependency between two constraints is recognized as an edge and represents the data flow. Meanwhile, it also updates the depth and out-degree of the node accordingly (Lines \ref{stmt:add_edge_start}-\ref{stmt:add_edge_end}). 

Next, the nodes in $G^{-1}$ without incoming edges are termed as the root set $L_{root}$, which are the source nodes to perform sorting (Line \ref{stmt:root}). The upper bound $s$ of partition size is set to $\left\lceil {|V|}/{k} \right\rceil$ (Line \ref{stmt:partition_size}). After, it goes into the process of partitioning  (Lines \ref{stmt:root_partition_start}-\ref{stmt:root_partition_end}).  Specifically, it traverses from a node in $L_{root}$ and explores a connected component recursively by function Rec\_Partition $C$ (Lines \ref{stmt:rec_partition_start}-\ref{stmt:rec_partition_end}) to add nodes into the current partition. When visiting children of $v$ recursively, the algorithm accesses each child in the order of their priorities based on the depth and out-degree, as discussed in the heuristic strategies (Lines \ref{stmt:explore_children_start}-\ref{stmt:explore_children_end}). Then node $v$ is assigned to the current partition $V_p$ once all its descendants, the nodes it depends on, are processed (Line \ref{stmt:add_to_partition}). If the size of $V_p$ reaches the threshold $s$, $V_p$ is inserted into $\mathcal{P}$ as a independent partition (Lines \ref{stmt:add_new_partition_start}-\ref{stmt:add_new_partition_end}). Finally, the partition $\mathcal{P}$ is returned. 

This traversal ensures that nodes are assigned to partitions in a way that respects the dependencies captured by the Rec-CDG while prioritizing child nodes based on depth and out-degree reduces shared variables between partitions.

%forked from the original scheduler

\section{Scheduler} \label{sec:scheduler}
In this section, we introduce the \emph{scheduler}, responsible for 
coordinating the executions for subcircuits based on a scalable pipeline model. 

\vspace{-3pt}
\subsection{Subcircuit Construction \& Execution}
Given a partition $\mathcal{P}=\{V_1, \ldots, V_k\}$, the scheduler should construct a subcircuit $F_i$ for each partition $V_i$, which is actually executable as shown in the \cref{fig:circuit_partition}. If the scale of the original circuit $\mathcal{F}$ is large, the resources, e.g., the memory, equipped may not afford its execution. With smaller subcircuits, the execution for which consumes much fewer resources, \system can first execute them separately and then combine the executions to produce the final proof. 
Essentially, it must ensure the combination of these subcircuits' executions is equivalent to the full circuit.

\paragraph{Subcircuit construction} Given a \zk circuit $\mathcal{F}$, the subcircuit $F_i$ corresponding to $V_i$ is defined as:
$$
F_i\left(I_p^i, I_s^i, \bigcup_{t=1}^{i-1}{S_{ti}}\right) = \left(O_p^i, O_s^i, \bigcup_{t=i+1}^{k}{S_{it}}\right)
$$
\noindent The inputs of $F_i$ encompasses three parts. The $I_p^i\subseteq I_p$ and $I_s^i\subseteq I_s$ are the subsets of public and secret inputs involved in $F_i$, respectively. Each $S_{ti}$ indicates the intermediate results to be transmitted from a previous subcircuit $F_t$ to $F_i$. In fact, $S_{ti}$ corresponds to the cross-partition edges in CDG $G$ between $V_t$ and $V_i$. 
Then,  $\bigcup_{t=1}^{i-1}{S_{ti}}$ merges all the inputs received from previous subcircuits, termed as the \emph{shared inputs}. Similarly,  $O_p^i$ and $O_s^i$ denote the public and secret outputs of $F_i$, maintaining the output structure of the original circuit, while $\bigcup_{t=i+1}^{k}{S_{it}}$, the \emph{shared outputs}, are going to be passed to succeeding subcircuits. In \system, public outputs can be exposed to the verifier, while secret outputs are only for internal use to the prover.

\paragraph{Subcircuit-wise execution} 
Given the set of $\{F_1, \ldots, F_k\}$ regarding $\{V_1, \ldots, V_k\}$, \system can execute them sequentially to generate the final proof. Some outputs propagate onward as shared variables during this process, connecting the subcircuits. The serializability of partition ensures that the current subcircuit $F_i$ can continue computation based on the shared variables from every proceeding subcircuit $F_t$ ($t<i$). Obviously, the combined logic of these subcircuits is equivalent to the full circuit.

However, this sequential execution model cannot utilize the advantage of modern multi-core chips. Actually, we can parallelize the execution of subcircuits if they have no dependency. For example, the subcircuit $F_1$ and $F_2$ in \cref{fig:circuit_partition} can be executed in parallel. \system adopts a \emph{execution DAG} (ExDAG) to capture dependencies between subcircuits. In an ExDAG, each node represents a subcircuit $F_i$, and $F_i$ points to $F_j$ iff %= if and only if
the counterpart partition $V_i$ is depended on by $V_j$. Initially, the subcircuits can be executed in parallel, not depending on others. As the execution proceeds, the edges between nodes (subcircuits) are moved and new subcircuits are available for parallel execution until all are finished.

Unfortunately, such an execution model still falls in the low CPU utilization as illustrated in \cref{fig:solve_vs_prove}. Therefore, \system proposes a scalable pipeline execution framework, achieving high resource utilization. 

\begin{figure}[t]
    \centering
    \includegraphics[width=0.47\textwidth]{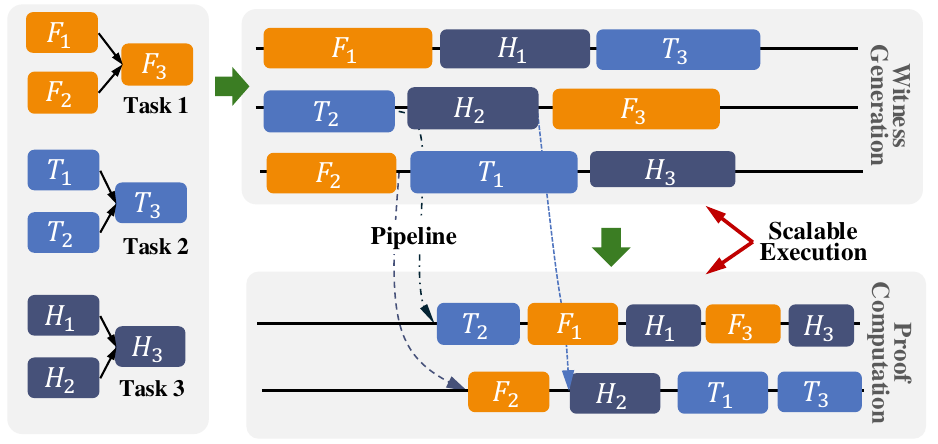}
    \caption{Pipelined and parallel execution of the Scheduler.}
    \label{fig:scheduler}
\end{figure}

% 5.2节，就是解耦了witness generation和proof computation，前者cpu利用率差，很慢；后者经过优化以后可以很好的利用cpu（具体看电路规模，大的时候可以有70~80%甚至100%，可以参考图1）；同一个子电路的p_c一定要等它的w_g做完；但不同子电路的完全独立；然后这里我们更在意的是源源不断的证明任务，每个任务task的电路是一样的（也就是子电路一样），只是输入不一样；这些任务之间w_g和p_c完全独立，想怎么并行怎么并行；

% 可配置的原因是，比如一个w_g要20s，p_c要4s，那么单纯的1:1的pipeline其实还是受限于w_g，所以要根据实际情况配置两者的比例，比如上面那个里子，就是5:1比较好;但这样会带来内存的问题，5个w_g带来的内存会很大，而通过partition，比如5-partition，把这5个w_g的内存变成相当于1个，这样双赢；

% 一个可能的想法是自适应的配置，这个和电路相关，目前不确定所以没做，可以作为一个讨论

% 下面的逻辑就是前面在说pipeline，中间说怎么做，最后举了个简单的例子，这个例子可以和figure5绑定，我专门弄了F5里的三个子电路的顺序和这个例子是一样的，就是1和2子电路可并行，3依赖于1,2

% 第4节的“可串行”其实是包含一定的并行可能的，所以其实某种意义上下面的pipeline里的调度算法是可以有一些深层考虑的，也是一个讨论的方向；现在就是单纯的没依赖了，拿出来准备调度，例子里只是一个可能的调度；目前没细想

%Upon obtaining the subcircuits and their corresponding execution order, the \textbf{Scheduler} orchestrates a \textit{configurable pipeline} to manage the execution of these subcircuits in response to the continuous stream of proof requests.

\subsection{Scalable Pipeline Execution}

\paragraph{Pipeline} 
To increase resource utilization with continuous proof generation requests, termed \emph{tasks}, \system introduces a pipeline design to handle tasks simultaneously. Each task can naturally be divided into two phases: witness generation (WG) and proof computation (PC) as described in \cref{subsec:zk}.  The execution for each subcircuit also goes through the two phases. A straightforward way is to decouple the WG and PC and allow two parallel phases to process different tasks. Once WG for the former task ends, the next task's WG is scheduled in parallel to the PC of the former task. 
For the pipeline to optimize resource utilization effectively, the precondition is that the latencies of the WG and PC phases are approximately equal. However, \cref{fig:solve_vs_prove} demonstrates that WG requires much more time than the PC in the current optimized {\zk}. Consequently, overall resource utilization still remains low.

\paragraph{Scalable execution} If the computational ability of each phase is scalable, \system can balance the latencies by delicately assigning resources to both phases. In this setting, the resource utilization can be maximized. To this end, the scheduler employs two groups of workers: \emph{solve workers} for WG and \emph{prove workers} for PC. With more workers devoted, the performance of the WG phase can be raised to a level comparable to the PC phase, and the number of both kinds of workers relies on concrete cases. 

With continuous tasks received, the scheduler obtains their ExDAGs and assigns the subcircuits that depend on no others to multiple solve works for parallel witness generation. As this phase for some subcircuits finishes, other subcircuits, depending on them as per the ExDAGs, become ready and are gradually scheduled for witness generation. Then, each subcircuit is delivered to a prove worker for proof computation. The PC for different subcircuits can proceed in parallel, as there are no further dependencies after the witness generation.

\begin{example}   
\Cref{fig:scheduler} illustrates how witness generation and proof computation phases can operate in parallel for three tasks: $F$, $T$, and $H$. Each of them is partitioned into three subcircuits again. Here, the number of solve and prove workers is 3 and 2, respectively.  Initially, the subcircuits $F_1$, $F_2$, and $T_2$ are assigned to three solve workers for witness generation, achieving an intra- and inter-task concurrency. This is because they do not depend on other subcircuits. 
Once WG for $F_2$ or $T_2$ ends, it is delivered to PC phase, handled by a prove worker. In the PC, all subcircuits can be executed concurrently in any arbitrary order, as the proof computation for each is fully independent. As we can see, all workers stay busy during the process, maximizing resource utilization. If the performance of the witness generation phase cannot catch up with the proof computation phase, \system can devote more skilled workers to achieve a new balance in \system's scalable framework.
\end{example}

Different hardware environments and circuit structures exhibit varying resource utilization and execution times for both phases. In some cases, WG may take significantly longer than PC, while in others, they may have close running time.   We should configure the number of both kinds of workers catering to the real workloads.

\begin{algorithm}[t]
\caption{Preprocessing for execution}
\label{alg:Subcircuit Construction}  % 添加label
\require{ \rm CDG $G = (V, E)$, partition $\mathcal{P} = \{V_1, \ldots, V_k\}$}

% 这里大写的F用在上面对原电路的定义里了
\ensure {\rm Subcircuits $F = \{F_1, \ldots, F_k\}$, public inputs $I_p$, secret inputs $I_s$, ExDAG $D = (\mathcal{V}, \mathcal{E})$  }

Initialize subcircuits set $F \gets \emptyset$ \\
Initialize constraints set $C_p \gets \emptyset$, input set $I \gets \emptyset$\\
Initialize shared variable sets $S_{ij} \gets \emptyset, \forall 1\le i\ne j \le k$  \\

Initialize an graph $D \gets (\mathcal{V}=\{F_1,\ldots, F_k\}, \mathcal{E}=\emptyset)$ \\

\For{\rm each partition  $V_i \in \mathcal{P}$} {
    \For{\rm each node $u \in V_i$} {
        $C_p \gets C_p \cup \{u\}$ \label{stmt:constraint_add}\\
        \For{\rm each $u$'s child node $v$ \label{stmt:process_child_start}} {
            \If{\( v \in V_j \) and \( i \neq j \)} {
            $S_{ij} \gets S_{ij} \cup \{u\}$  \\   
                $\mathcal{E}\gets \mathcal{E}\cup\{(F_i, F_j)\}$ \label{stmt:process_child_end}
            }
        }
    }
    $I \gets$ Get\_Input $(C_p, I_p, I_s)$ \label{stmt:construct_subcircuit_start}\\
    $F_i \gets$Create\_Subcircuit$(I, C_p, \cup_{t=1}^{i-1}S_{ti}, \cup_{t=i+1}^{k}S_{it})$ \label{stmt:construct_subcircuit_end}\\
    $F\gets F\cup F_i$ \\
    $C_p, I \gets \emptyset$\\
}

\Return{$F, D$}
\end{algorithm}

\subsection{Algorithm}

\Cref{alg:Subcircuit Construction} outlines the preprocessing for the scalable pipeline execution, including the construction of subcircuits and the ExDAG. Each partition  $V_i$ identifies the necessary inputs, constraints, and shared variables to form the corresponding subcircuit $F_i$. Specifically, the constraints associated with the nodes in the partition $V_i$ are added to the constraint set (Line \ref{stmt:constraint_add}). When a node $u\in V_i$ points to a node $u$ in a different partition $V_j$ ($j > i$), this establishes a cross-partition dependency. The output of the node $v$ is recorded as a shared variable between $F_i$ and $F_j$, and then a edge is added into $D$
(Lines \ref{stmt:process_child_start}-\ref{stmt:process_child_end}). Once all nodes in partition $V_i$ are processed, the algorithm gathers the necessary inputs, including any shared variables from prior partitions, and constructs the subcircuit $F_i$ (Lines \ref{stmt:construct_subcircuit_start}-\ref{stmt:construct_subcircuit_end}).

\begin{algorithm}[t]
\caption{Scalable Pipeline Execution}
\label{alg:pipeline}

Initialize an empty solve queue \( Q_{\text{solve}} \) and an empty prove queue \( Q_{\text{prove}} \) \label{stmt:pipeline_init_start}\\
Initialize task pool $\mathcal{P}_{task}\gets \emptyset$ \label{stmt:pipeline_init_end}

\upon{\rm receive task $R$ \label{stmt:task_receive_start}}{
    $D\gets Get\_ExDAG(R)$\\
    Add $D$ to  $\mathcal{P}_{task}$ \label{stmt:task_receive_end}\\
}

\vspace{3pt}

\tcp{Run on a separate thread} 
\While{\rm true \label{stmt:schedule_start}} {
    $F_i\gets Next\_Ready\_Subcircuit(\mathcal{P}_{task})$  \tcp{All dependencies are resolved}
    $Q_{solve}.Push(F_i)$ \label{stmt:schedule_end}\\ 
}
\vspace{3pt}

\tcp{Run on each solve worker in parallel}  
\While{\rm true \label{stmt:solve_state_start}} {
    $F_i\gets Q_{solve}.Pop()$\\
    $Witness\_Gen(F_i)$\\
    
    $C \gets Get\_Dependent\_Circuits(\mathcal{P}_{task}, F_i)$ \tcp{Get all circuits depending on $F_i$}
    \For{each subcircuit \( F_j \in C\)}{
        $Share\_Variables(F_i, F_j)$ \tcp{Share the output with dependent subcircuits}
    }

    $Q_{prove}.Push(F_i)$ \label{stmt:solve_state_end}
}
\vspace{3pt}

\tcp{Run on each prove worker in parallel}
\While{\rm true \label{stmt:prove_stage_start}} {
    $F_i\gets Q_{prove}.Pop()$\\
    $Proof\_Com(F_i)$\\
    $Finish(F_i, \mathcal{P}_{task})$\label{stmt:prove_stage_end}
}
\end{algorithm}

As the subcircuits and execution DAG are constructed, we can present the framework of our scalable pipeline model as elaborated in \Cref{alg:pipeline}. Specifically, it goes through the following stages:
\begin{itemize}[leftmargin=*]
    \item \textbf{Initialization (Lines \ref{stmt:pipeline_init_start}-\ref{stmt:pipeline_init_end}):}  
    Initially, the algorithm prepares two empty queues: $Q_{\text{solve}}$ for subcircuits ready for witness generation, and $Q_{\text{prove}}$ for subcircuits awaiting proof generation. Besides, a task pool $\mathcal{P}_{task}$ is used to store all tasks' subcircuits. 

    \item \textbf{Task receiving (Lines \ref{stmt:task_receive_start}-\ref{stmt:task_receive_end}):} Upon receiving a task to generate a proof, the execution DAG for this task is retrieved and added to the task pool $\mathcal{P}_{task}$.

    \item \textbf{Scheduling (Lines \ref{stmt:schedule_start}-\ref{stmt:schedule_end}):}  
    The scheduler continuously checks for any subcircuit $F_i$ ready for witness generation and adds it to queue $Q_{\text{solve}}$ to be consumed by available solve workers. A subcircuit is ready if all its dependencies in the DAG are resolved. 
    
    \item \textbf{Solve phase (Lines \ref{stmt:solve_state_start}-\ref{stmt:solve_state_end}):}  
    Each solve worker continuously acquires subcircuit $F_i$ from $Q_{\text{solve}}$ and performs witness generation. After that, the worker passes the shared outputs of $F_i$ to other subcircuits that depend on it to trigger succeeding execution. Then, $F_i$ is inserted into the prove queue $Q_{\text{prove}}$ for proof computation.

    \item \textbf{Prove phase (Lines \ref{stmt:prove_stage_start}-\ref{stmt:prove_stage_end}):}  
    Each prove worker separately performs proof computation for subcircuits retrieved from $Q_{\text{prove}}$ and then marks $F_i$ as completed.
\end{itemize}

% 和partition结合的好处和理由

\subsection{Discussion}

The scalable pipeline increases overall memory usage as more subcircuits are executed on different workers simultaneously. Therefore, the partitioning approach should break the circuit into suitable and manageable subcircuits, promising the memory required by all subcircuits to be executed in parallel is under limit. By tuning the factor $k$, the number of partitions, it allows \system to better control memory consumption while effectively utilizing resources. Building on the circuit partition and scalable pipeline, \system can maximize the throughput and be able to handle larger \zk workloads efficiently.

\system currently relies on manual configuration of the ratio between solve and prove workers to adapt to different scenarios. While this approach provides flexibility, a more promising direction is to tune these parameters in an adaptive manner. Specifically, as tasks are processed, \system should adjust the allocation of solve and prove workers based on real-time statistics of witness generation and proof computation phases. \system can collect this information by continuously evaluating the latencies and resource utilization during execution. This dynamic adjustment would optimize performance and resource utilization, especially in environments with varying workloads. 

\section{Evaluation} \label{sec:evaluation}
In this section, we evaluate the efficiency of  \system and compare it with the current baseline.

\subsection{Experimental setup}\label{6.1}

\paragraph{Implementation} We built Yoimiya based on the Gnark library\footnote{\url{https://github.com/ConsenSys/gnark}}, a high-performance \zk framework written in Go, which provides a high-level API for designing cryptographic circuits. In our implementation, we adopt the Rank-1 Constraint System (R1CS) \cite{parno2016pinocchio} to represent the constraints. The bilinear map used in Gnark is instantiated using the BN254 curve\footnote{\url{https://github.com/Consensys/gnark-crypto}}, which offers approximately 100 bits of security. This curve's pairing operations are also supported in Solidity, the programming language used for Ethereum smart contracts.

\paragraph{Workloads} Our circuit is constructed using Gnark and represents a zero-knowledge proof for a simple linear recursive sequence, defined as $F_n = \alpha F_{n-1} + \beta F_{n-2}$. This allows us to prove that the $n$-th term in the sequence is equal to a specified value. We can vary the constraints in the corresponding constraint system by adjusting the number of iterations. The constraint number of the circuit used in our experiment is up to 60 million.

\paragraph{Metrics}
We assess the effectiveness and efficiency of \system from the following  metrics:
\begin{itemize}[leftmargin=*]
    \item Prove Memory: the memory required during the proof generation, including the memory consumed by witness generation and proof computation.

    \item Total Memory: the overall memory usage, counting the prove memory and additional memory used by other components, i.e., the constraint system and the key management system.
    \item Prove Time: the total time to generate all proofs, excluding the one-time setup of the circuit and the verification phases on the client sides.
    \item CPU Usage: the CPU utilization over time during the proof generation process.

\end{itemize}

\paragraph{Testbed}  We performed experiments on a server equipped with an Intel Xeon Gold-6330 2.00GHz CPU with 56 cores, 112 threads, and 500GB RAM, running Ubuntu 20.04.2 LTS. Building on the resources available and the real workloads, we set the number of solve and prove workers to 4 and 1, respectively, which appropriately balances the latency for both phases in our experiments. We run each set of experiments multiple times and take the average.

%To simplify the experimental setup, we set the number of prove workers to 1, treating multiple instances of proof computation as independent tasks. The number of solve workers in our experiments is up to 4.

\subsection{Performance of Circuit Partitioning}
\label{6.2}
 We evaluate the effectiveness of circuit partitioning on memory usage and proof generation time.

% \begin{figure}[t]
% \centering
% \includegraphics[width=0.48\textwidth]{figures/graph_size_performace.pdf}
% \caption{Performance of the normal and 2-partition approaches under different loop sizes.}
% \label{fig:graph_size_performance}
% \end{figure}
% 第一部分实验就是针对Partition
% 1. 改变电路规模（电路是循环，所以调整循环数即可），固定partition数为2，说明划分算法在电路规模上的可扩展性
% 2. 改变分区数为2~5，固定电路，说明划分算法在分区数上的可扩展性

\begin{figure}[t]
    \centering
    \captionsetup[subfigure]{skip=0pt}
    \begin{subfigure}[b]{0.15\textwidth}
        \centering
        \includegraphics[width=\textwidth]{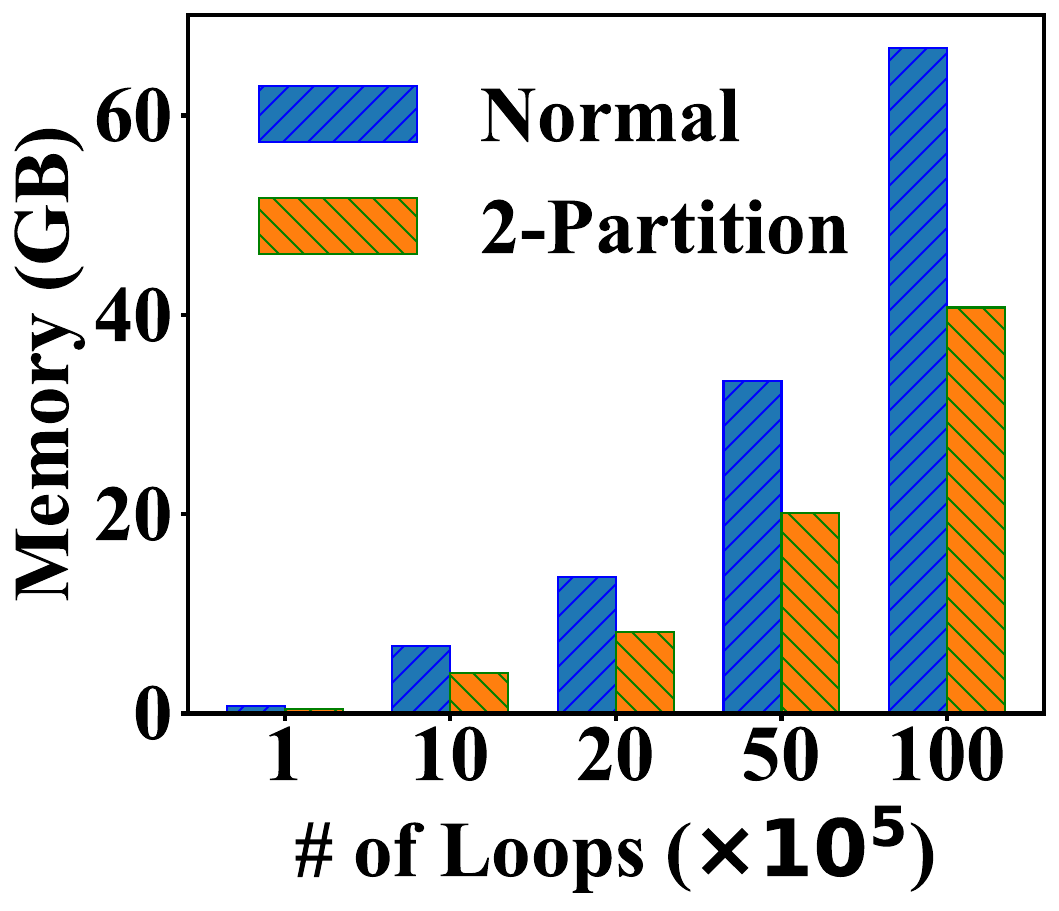}
        \captionsetup{justification=centering}  % Center the caption
        \caption{Total Memory}
        \label{fig:graph_size_performance_a}
    \end{subfigure}
    \hfill
    \begin{subfigure}[b]{0.15\textwidth}
        \centering
        \includegraphics[width=\textwidth]{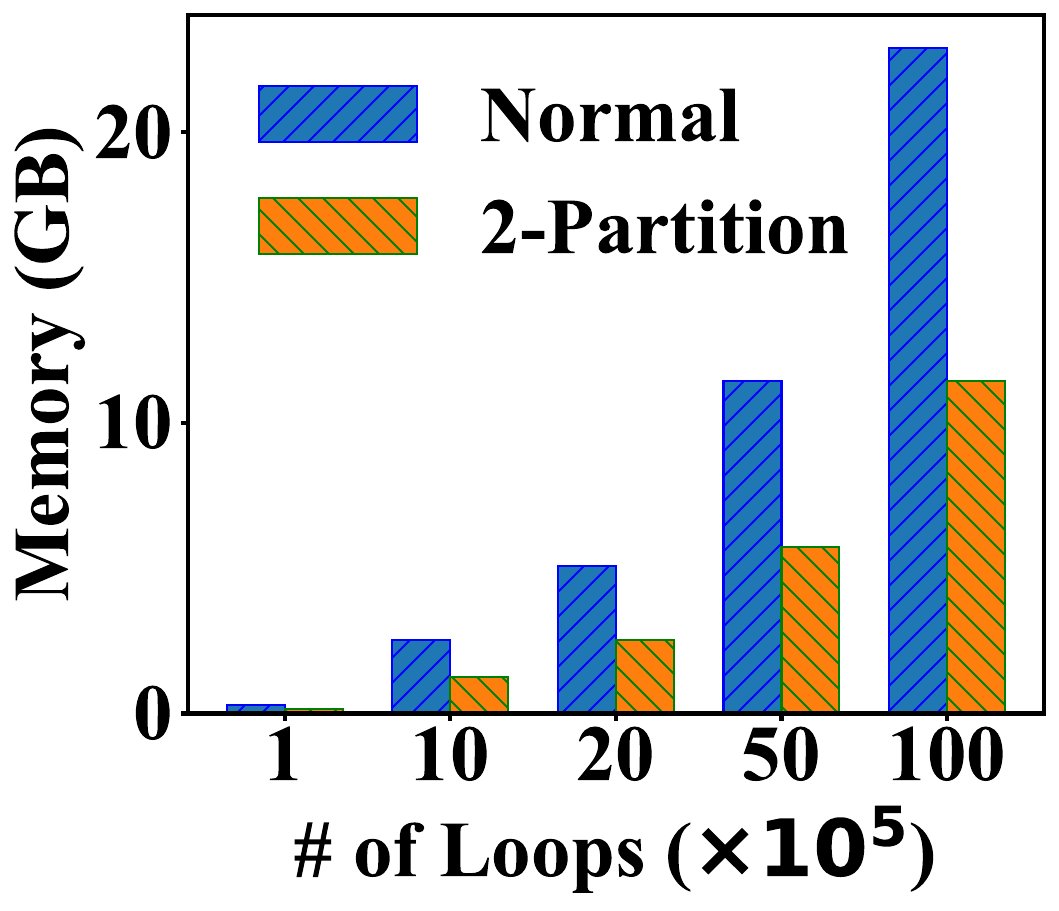}
        \captionsetup{justification=centering}  % Center the caption
        \caption{Prove Memory}
        \label{fig:graph_size_performance_b}
    \end{subfigure}
    \hfill
    \begin{subfigure}[b]{0.15\textwidth}
        \centering
        \includegraphics[width=\textwidth]{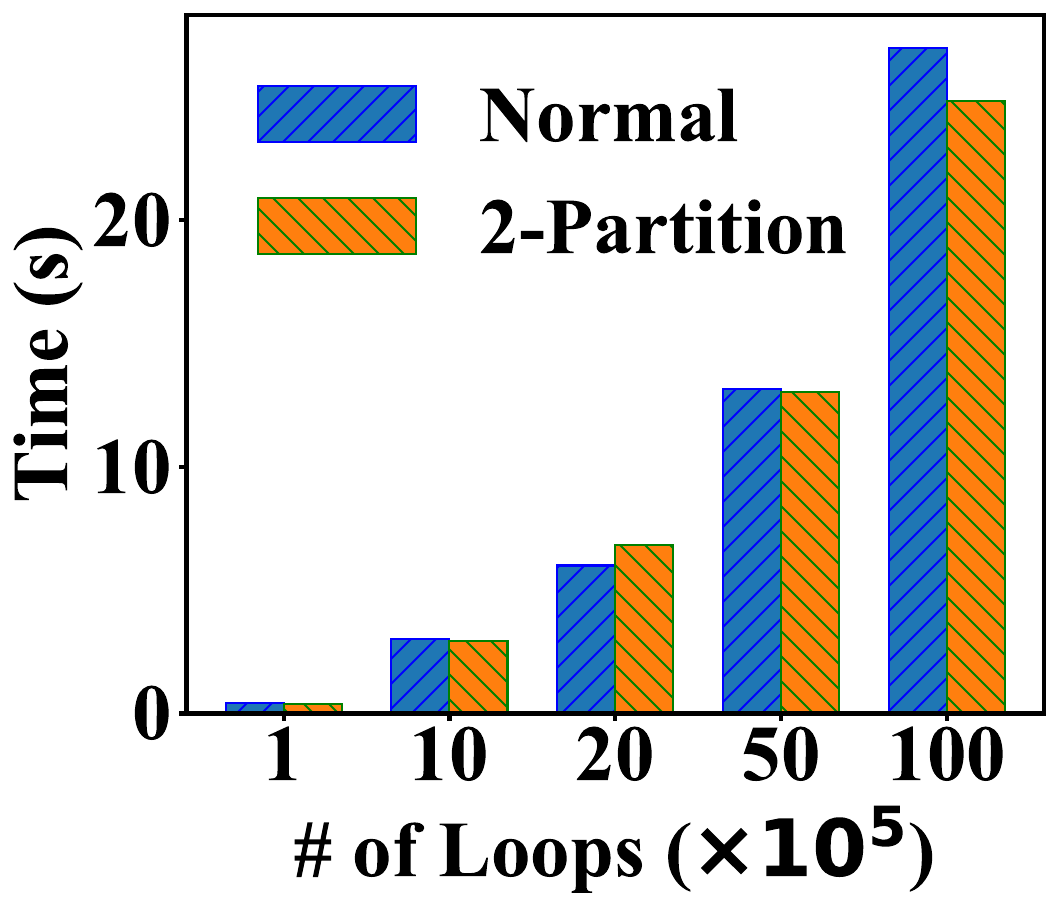}
        \captionsetup{justification=centering}  % Center the caption
        \caption{Prove Time}
        \label{fig:graph_size_performance_c}
    \end{subfigure}
    \caption{Performance of the normal and 2-partition approaches under different loop sizes.}
    \label{fig:performance_comparison}
\end{figure}

% 实验1：circuit(graph)_size_partition，结果就是内存要下去，这里pk没堆叠，所以total和prove都是明显下降，然后时间要尽可能一样，就是划分后跑2个和原来跑一个的时间差不多
\paragraph{Circuit size}
We first assessed the performance of our partitioning approach with varying circuit sizes, which is controlled by the loop count, as shown in \cref{fig:performance_comparison}. Meanwhile, the number of partitions is fixed to 2, and the loop count is up to 10 million. \crefrange{fig:graph_size_performance_a}{fig:graph_size_performance_b} showcases that the circuit partitioning significantly reduces total and prove memory consumption. Moreover,  the reduction becomes more pronounced as the circuit size increases. For example, the prove memory consumption with 2-partition is 59\% of the normal case when loop count is 100K while it decreases to 51\% with 10 million loops. 
Importantly, \cref{fig:graph_size_performance_c} confirms that the gap of total proof generation between the two approaches remains slight (up to 13\%), not sacrificing too much latency. This is critical for the scalable pipeline design, as the partitioning approach will not significantly raise the latency of circuit execution.

%the scalability of our partitioning approach by varying the size of the circuits, controlled through the loop count, while keeping the number of partitions fixed at 2. 

% First, we evaluate the impact of circuit partitioning on memory usage and proof generation time for different circuit sizes(by adjusting the loop number). To control the variables, we set the number of partitions to 2. We mainly tested three indicators: \textbf{Total Memory}, \textbf{Prove Memory}, \textbf{Prove Time}. \textbf{Total Memory} refers to the circuit-specific memory required for each subcircuit, such as the proving key and constraint system, as well as the memory used during the actual proof generation process(denoted as \textbf{Prove Memory}). \textbf{Prove time} refers to the time required to generate the zero knowledge proof corresponding to the generated circuit (or all sub circuits).

\begin{figure}[t]
    \centering
    \captionsetup[subfigure]{skip=0pt}
    \begin{subfigure}[b]{0.22\textwidth}  % Adjust width to balance in 2x2 layout
        \centering
        \includegraphics[width=\textwidth]{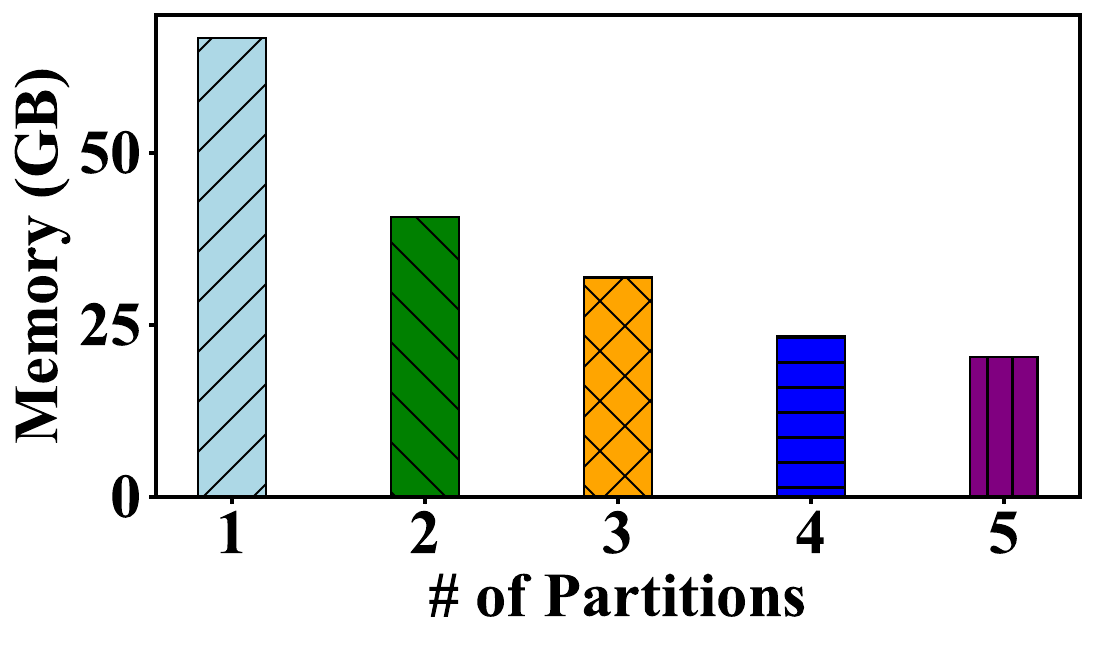}
        \captionsetup{justification=centering}  % Center the caption
        \caption{Total Memory}
        \label{fig:k_partition_performance_a}
    \end{subfigure}
    \hspace{0.01\textwidth}  % Adjust horizontal spacing between subfigures
    \begin{subfigure}[b]{0.22\textwidth}  % Adjust width to balance in 2x2 layout
        \centering
        \includegraphics[width=\textwidth]{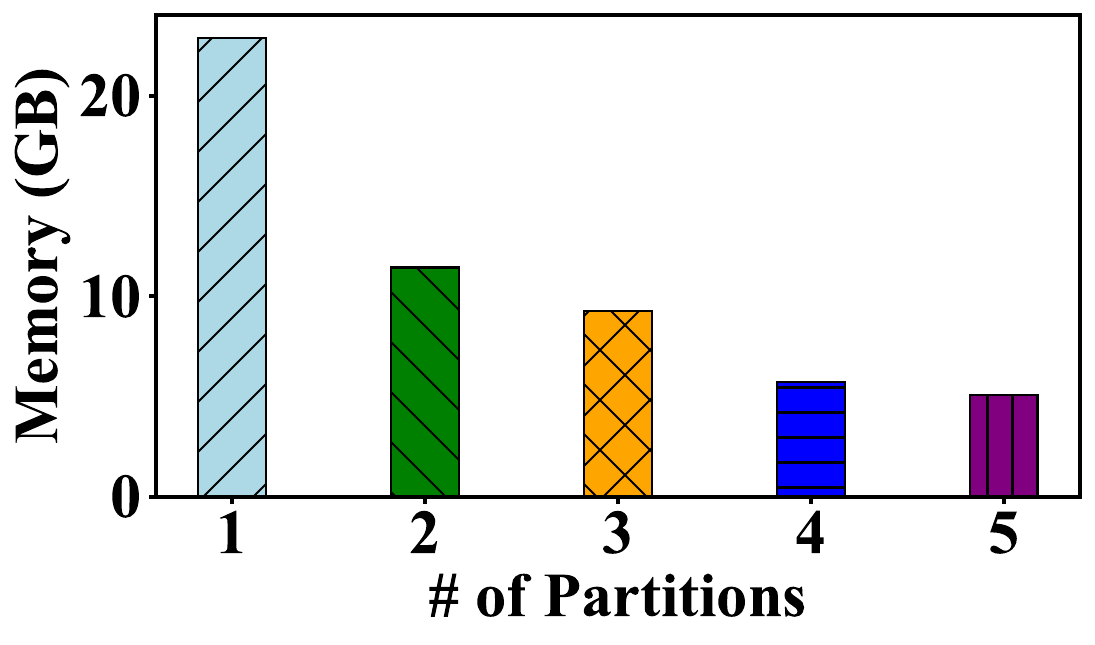}
        \captionsetup{justification=centering}  % Center the caption
        \caption{Prove Memory}
        \label{fig:k_partition_performance_b}
    \end{subfigure}

    \vspace{0.1em}

    \begin{subfigure}[b]{0.22\textwidth}  % Adjust width to balance in 2x2 layout
        \centering
        \includegraphics[width=\textwidth]{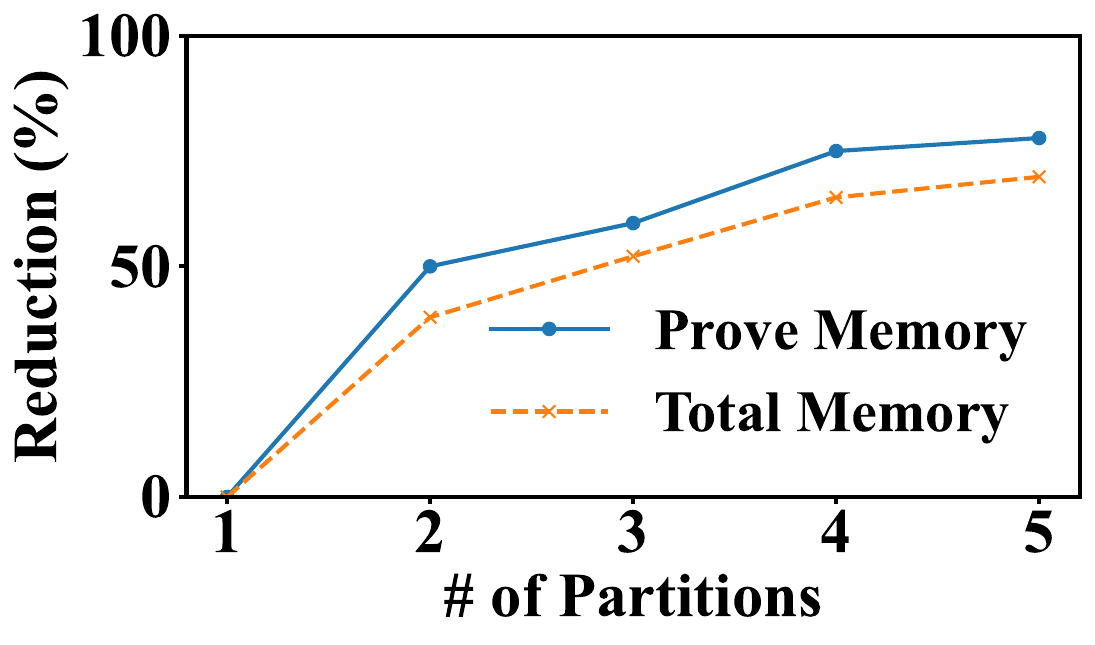}
        \captionsetup{justification=centering}  % Center the caption
        \caption{Memory Reduction}
        \label{fig:k_partition_performance_c}
    \end{subfigure}
    \hspace{0.01\textwidth}  % Adjust horizontal spacing between subfigures
    \begin{subfigure}[b]{0.22\textwidth}  % Adjust width to balance in 2x2 layout
        \centering
        \includegraphics[width=\textwidth]{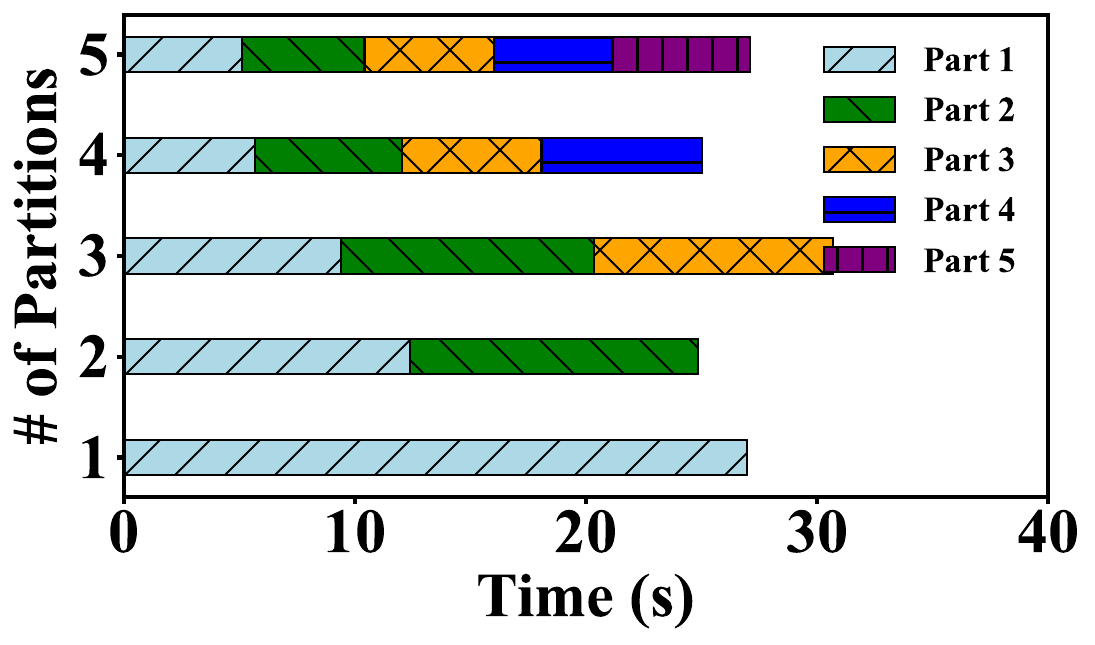}
        \captionsetup{justification=centering}  % Center the caption
        \caption{Prove Time}
        \label{fig:k_partition_performance_d}
    \end{subfigure}

    \caption{Performance for the normal and partitioned approaches across different partition numbers.}
    \label{fig:performance_comparison}
\end{figure}

% \begin{figure}[t]
% \centering
% \includegraphics[width=0.48\textwidth, trim={0 0 0 0}, clip]{figures/k_partition_performance.pdf}
% \caption{Performance for the normal and partitioned approaches across different partition numbers.}
% \label{fig:k_partition_performance}
% \end{figure}

% 实验二，partition_number_performance，就是改变分区数，内存和时间很直观，就是时间不怎么变，内存能下来；给了一个内存下降的比例（尽可能趋近反比例函数），凑图的；然后给出了一个不同分区占用的时间；
% 我这里还有setup的实验数据，也是时间和占比;
\paragraph{Partition number}
Next, we tested the partitioning approach with varying numbers of partitions, ranging from 1 to 5, while the constraint number is 60 million. A partition count of 1 stands for the normal case in previous tests. \crefrange{fig:k_partition_performance_a}{fig:k_partition_performance_b} show a near-linear reduction in both total memory and prove memory with a growth of partition number. In particular, the total memory consumption is reduced from 66.7 GB to 20.4 GB when the partition number reaches 5.  This trend is further confirmed in \cref{fig:k_partition_performance_c}, which presents the reduction in memory usage over partition number. Lastly, \cref{fig:k_partition_performance_d} breaks down the proof generation time for each partition. As expected, the time required for each partition goes from 27s to 6s as the number of partitions increases. This demonstrates that our partitioning strategy effectively reduces memory usage while maintaining stable execution time, establishing the foundation of the scalable pipeline execution.

% Next, we evaluate the effect of increasing the number of partitions on memory usage and proof generation time. In this experiment, we tested multiple partition configurations ranging from 1 to 5 subcircuits, where a partition count of 1 represents the standard (unpartitioned) execution. As seen in \Cref{fig:k_partition_performance_a} and \Cref{fig:k_partition_performance_b}, increasing the number of partitions leads to a nearly linear reduction in both total memory and prove memory. \Cref{fig:k_partition_performance_c} provides a more detailed view of this reduction, showing the percentage decrease in memory usage, which grows progressively with partitioning, illustrating the scalability of the partitioning approach. Lastly, \Cref{fig:k_partition_performance_d} breaks down the total proof generation time for each subcircuit, where individual circuits are denoted as $Part_i$. The time for each part decreases as the number of partitions increases, ensuring that the overall proof generation remains efficient while memory usage is significantly reduced.

% 第二部分实验，就是Pipeline以及pipeline+partition的效果
\subsection{Performance of Scalable Pipeline Execution}
\label{6.3}
% 一个电路，电路是实验一中最大的电路6000w约束，然后m:n的pipeline设置简化为了m:1，因为每个m：1看成一个独立的个体；然后测内存的时间多个子电路的pk是堆叠的，所以在这个实验里total的效果会比prove差一点；然后用20个证明请求来说明多任务
In this experiment, we evaluated the performance of our scalable pipeline framework towards continuous tasks.   To simulate such continuous tasks, we produce 20 tasks with approximately 60 million constraints and feed them into \system gradually. 
We want to test \system with varying rates between the number of solve and prove workers. Therefore, we fix one prove worker and vary the number of solve workers, as the witness generation requires much more time than proof computation.

\begin{figure}[t]
    \centering
    \captionsetup[subfigure]{skip=0pt}
    \begin{subfigure}[b]{0.15\textwidth}
        \centering
        \includegraphics[width=\textwidth]{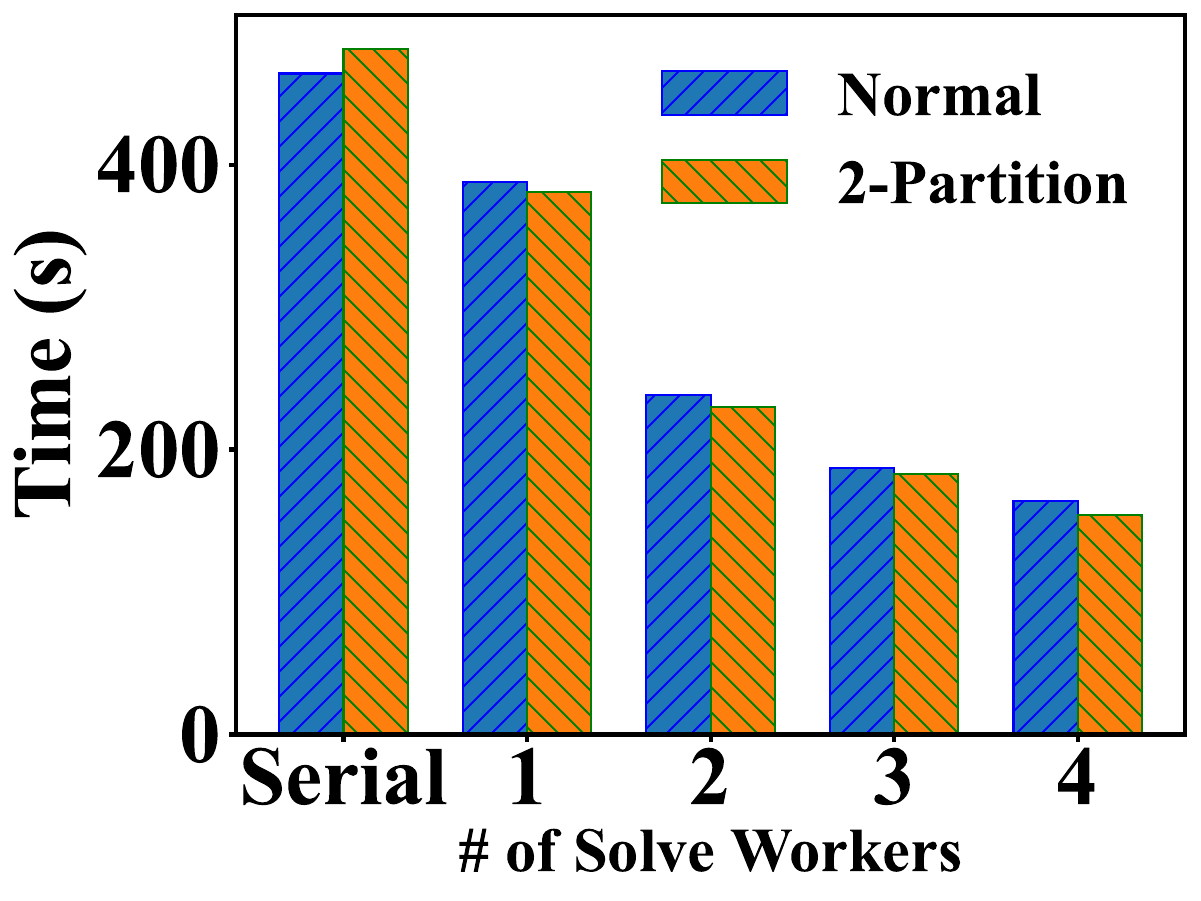}
        \captionsetup{justification=centering}  % Center the caption
        \caption{Time Cost}
        \label{fig:pipeline_solve_performance_a}
    \end{subfigure}
    \hfill
    \begin{subfigure}[b]{0.15\textwidth}
        \centering
        \includegraphics[width=\textwidth]{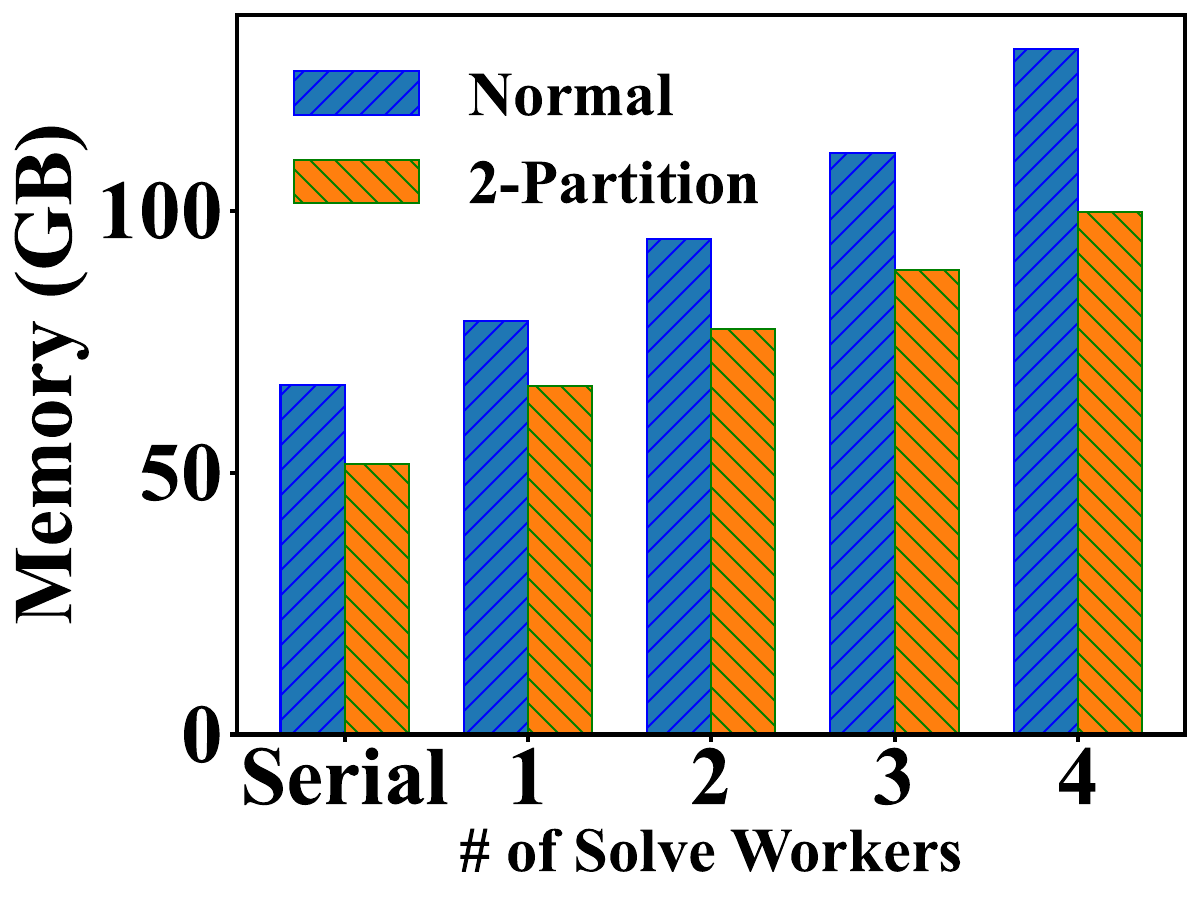}
        \captionsetup{justification=centering}  % Center the caption
        \caption{Total Memory}
        \label{fig:pipeline_solve_performance_b}
    \end{subfigure}
    \hfill
    \begin{subfigure}[b]{0.15\textwidth}
        \centering
        \includegraphics[width=\textwidth]{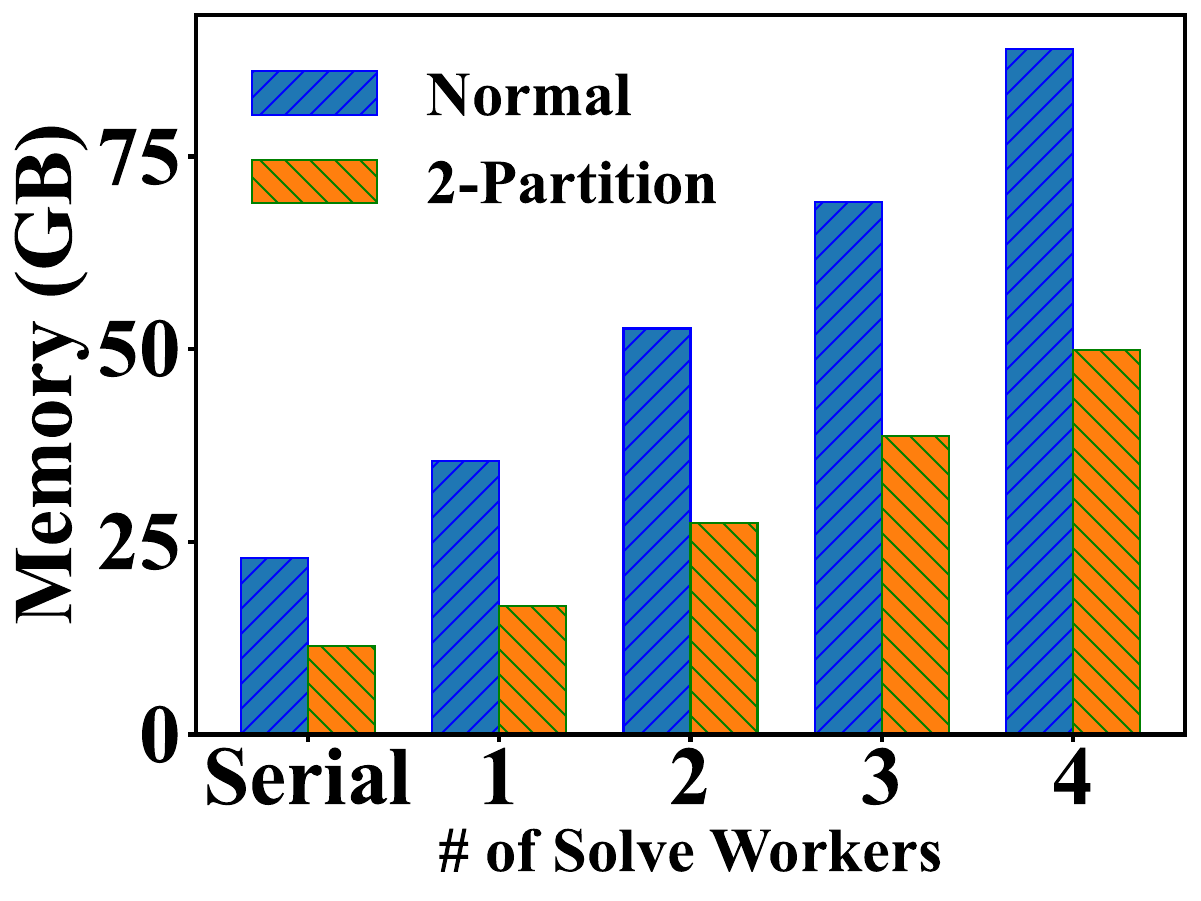}
        \captionsetup{justification=centering}  % Center the caption
        \caption{Prove Memory}
        \label{fig:pipeline_solve_performance_c}
    \end{subfigure}
    \caption{Performance for different numbers of solve workers across the normal and 2-partition approaches. }
    \label{fig:pipeline_solve_performance}
\end{figure}

\begin{figure}[t]
    \centering
    \includegraphics[width=0.46\textwidth]{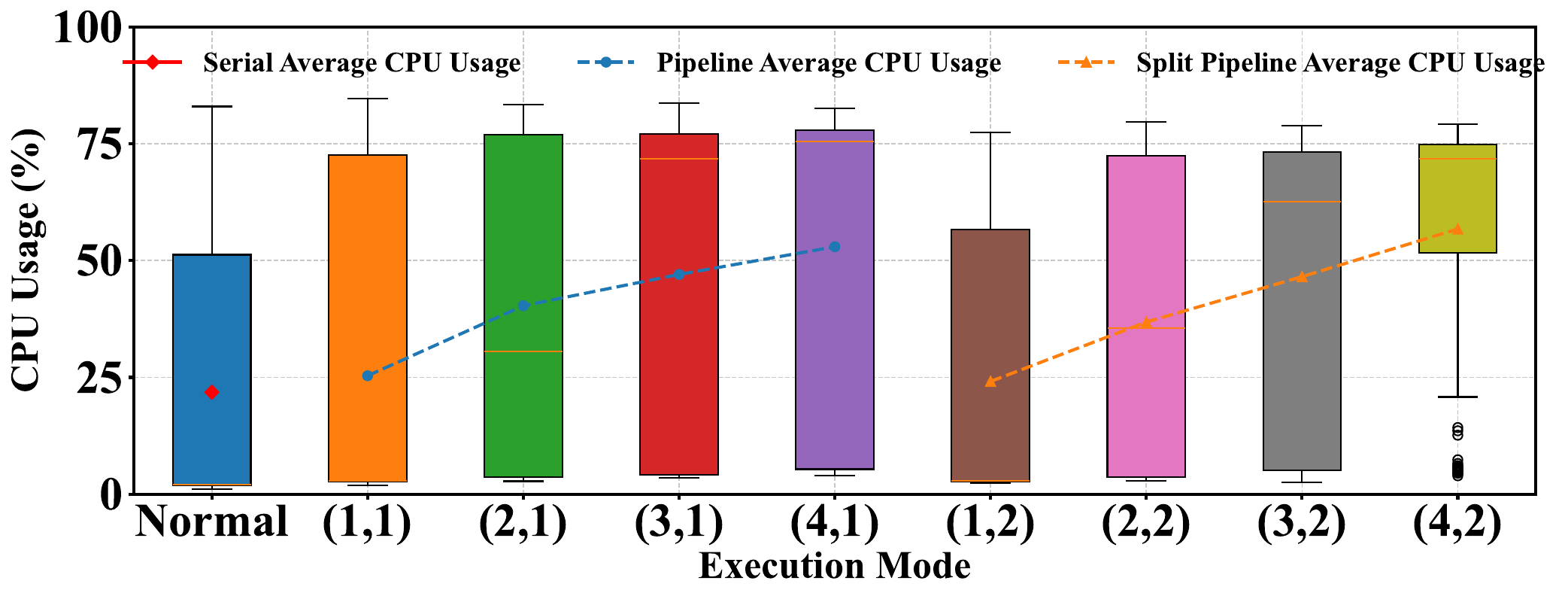}
    \caption{CPU usage distribution for different execution modes. Normal:  serial execution of all tasks; $(s,p)$: the number of solve workers is set to $s$, and the number of partitions is set to $p$. If $p=1$, it means no partitioning is applied. } 
    \label{fig:pipeline_solve_cpu_performance}
    
\end{figure}

\begin{figure}[t]
    \centering
    \includegraphics[width=0.48\textwidth]{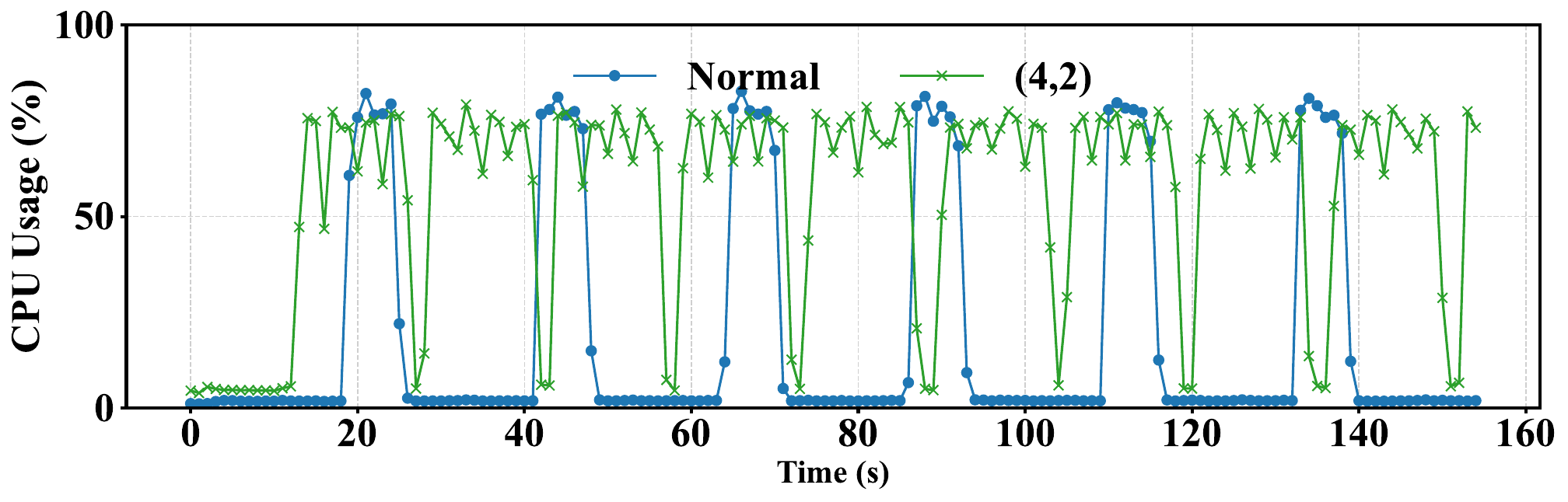}
    \caption{CPU usage with continuous tasks over time for different execution modes. } 
    \label{fig:pipeline_solve_cpu_performance}
\end{figure}

%To simplify the experimental setup, we set the number of prove workers to 1, treating multiple instances of proof computation as independent tasks. We fixed the number of tasks to 20, each involving circuits with approximately 60 million constraints, which is same as the largest circuit size used in the previous experiment.

% todo 没有说明pk的堆叠，但是否有必要说明？

% 首先测试的是pipeline，然后给出了pipeline+2-partition，因为要控制变量，这里的变量是witness generation的并行数0-4，要说明的就是1. 内存越来越大；2. 利用率越来越高；3. 时间越来越快；4. 用了2-partition的会比不用的内存小，时间、利用率差不多

\paragraph{Solve worker number}
First, we test the impact of solve worker number on the performance with 2 partitions for each circuit as reported in \cref{fig:pipeline_solve_performance}.
As illustrated in \cref{fig:pipeline_solve_performance_a}, increasing the number of solve workers significantly reduces the total time required to complete all tasks from 464s to 164s. This is because a prove worker outperforms a solve worker, so more solve workers must be devoted to increasing the parallelism, balancing their latency.  Note that when the performance of witness generation has aligned with proof computation, there is no need to scale solve workers. 
\crefrange{fig:pipeline_solve_performance_b}{fig:pipeline_solve_performance_c} reveals that as the number of solve workers increases, so does memory consumption. This because running multiple witness generation instances simultaneously is memory-intensive. \system partitions each circuit into more subcircuits, reducing the memory consumption while maintaining the parallelism between solve workers. In \cref{fig:pipeline_solve_performance_c}, the prove memory consumption for the 2-partition case is only 56.1\% of the single-partition case, lowering the memory barrier of the system.

 %However, \crefrange{fig:pipeline_solve_performance_b}{fig:pipeline_solve_performance_c} reveal a tradeoff with memory usage to be considered in practice.
%As shown in \Cref{fig:pipeline_solve_performance}. This combination reduced memory usage significantly while maintaining nearly identical proof generation times compared to the non-partitioned pipeline. This demonstrates that partitioning is an effective strategy to alleviate memory overhead, ensuring that memory consumption remains within manageable limits without compromising execution speed. 

\paragraph{Varying combination}
In addition, \cref{fig:pipeline_solve_cpu_performance} shows the CPU utilization of \system with various parameters. As the parallelism of the witness generation phase increases (with more solve workers), the CPU engagement improves significantly up to 52.9\% for the parameter (4,1) while the (1,1) case only exhibits 21.8\% CPU utilization. Additionally, the growth of partition number has slightly impacted the overall CPU utilization, which is 52.9\% and 56.7\% for (4,1) and (4,2). However, the partition approach can help to decrease the memory requirement significantly in our scalable pipeline framework as explained above. \cref{fig:pipeline_solve_cpu_performance} further elaborates the CPU usage of one typical case (4,2) in \cref{fig:pipeline_solve_cpu_performance} and the normal case over time. As we can see, the CPU usage for case (4,2) maintains over 60\%, mostly due to the scalable pipeline execution. The CPU usage for (4,2) also periodically falls low due to the sequential preparation phases during proof computation, which temporarily limits resource utilization. However, this period of case (4,2) is much shorter than the normal case.

\begin{figure}[t]
    \centering
    \captionsetup[subfigure]{skip=0pt}
    \begin{subfigure}[b]{0.22\textwidth}  % Adjust width to balance in 2x2 layout
        \centering
        \includegraphics[width=\textwidth]{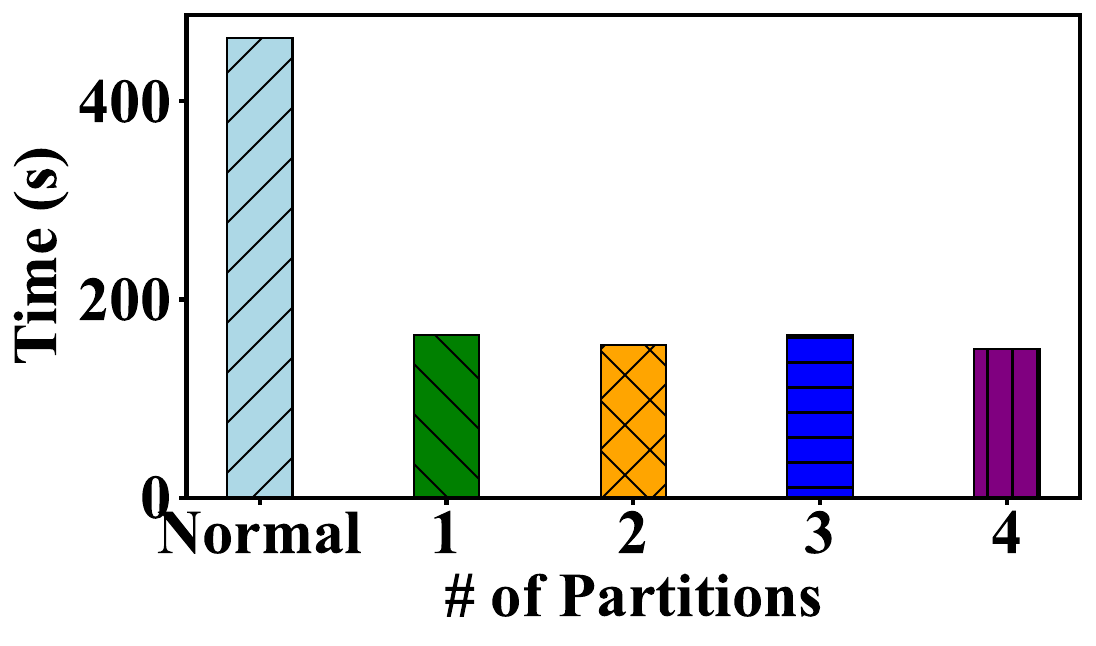}
        \captionsetup{justification=centering}  % Center the caption
        \caption{Prove Time}
        \label{fig:pipeline_partition_performance_a}
    \end{subfigure}
    \hspace{0.01\textwidth}  % Adjust horizontal spacing between subfigures
    \begin{subfigure}[b]{0.22\textwidth}  % Adjust width to balance in 2x2 layout
        \centering
        \includegraphics[width=\textwidth]{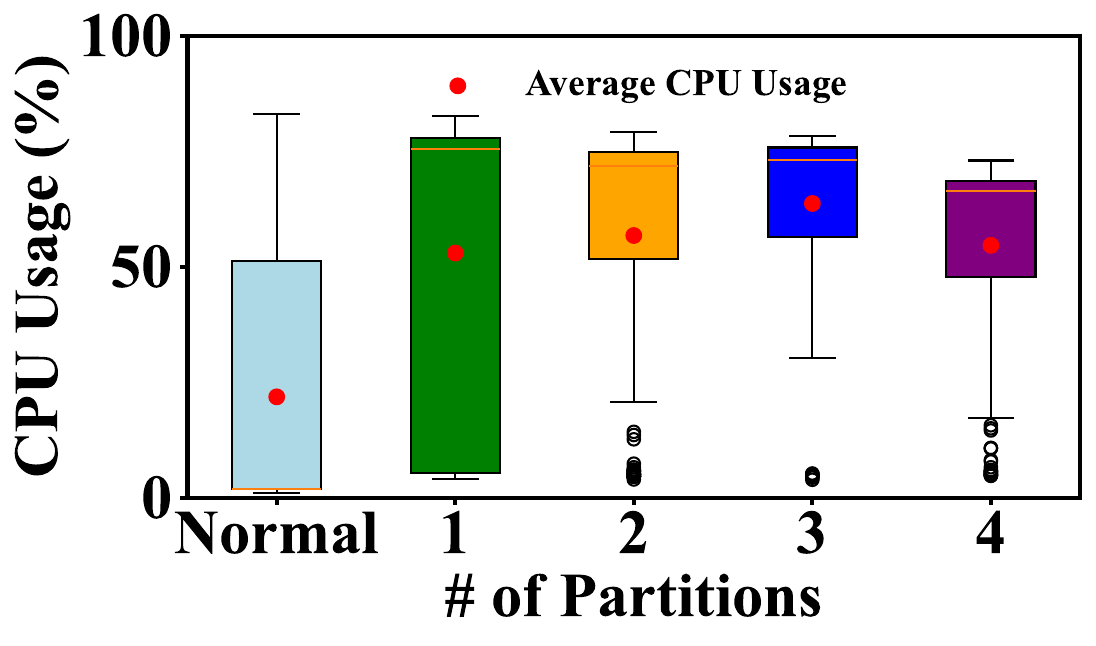}
        \captionsetup{justification=centering}  % Center the caption
        \caption{CPU Usage Distribution}
        \label{fig:pipeline_partition_performance_b}
    \end{subfigure}

    \vspace{0.1em}

    \begin{subfigure}[b]{0.22\textwidth}  % Adjust width to balance in 2x2 layout
        \centering
        \includegraphics[width=\textwidth]{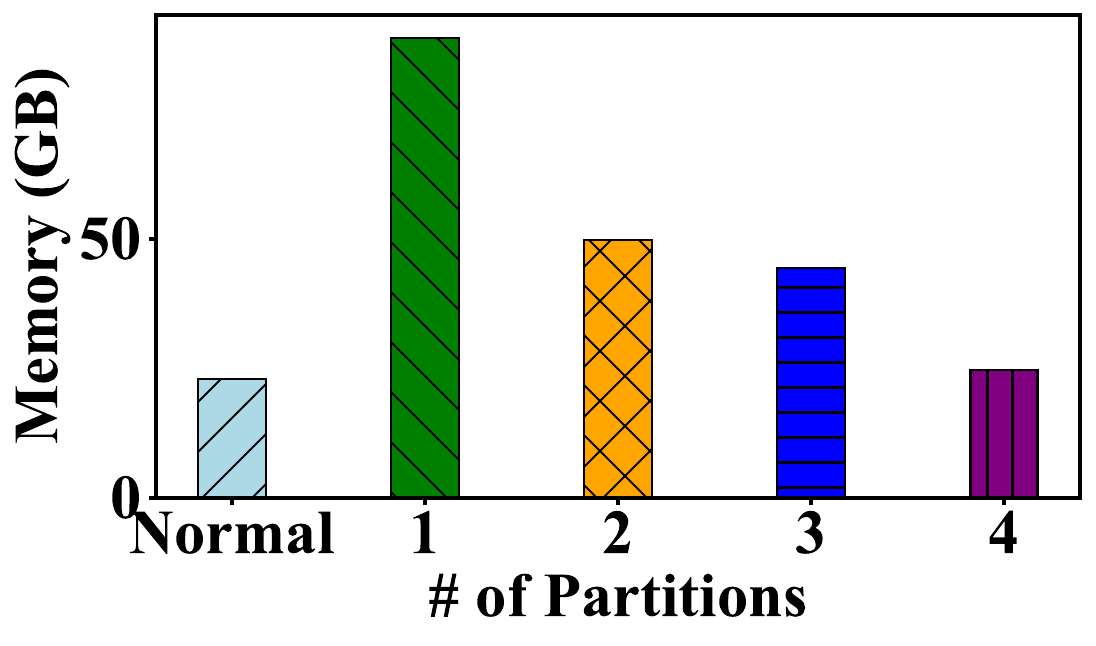}
        \captionsetup{justification=centering}  % Center the caption
        \caption{Prove Memory}
        \label{fig:pipeline_partition_performance_c}
    \end{subfigure}
    \hspace{0.01\textwidth}  % Adjust horizontal spacing between subfigures
    \begin{subfigure}[b]{0.22\textwidth}  % Adjust width to balance in 2x2 layout
        \centering
        \includegraphics[width=\textwidth]{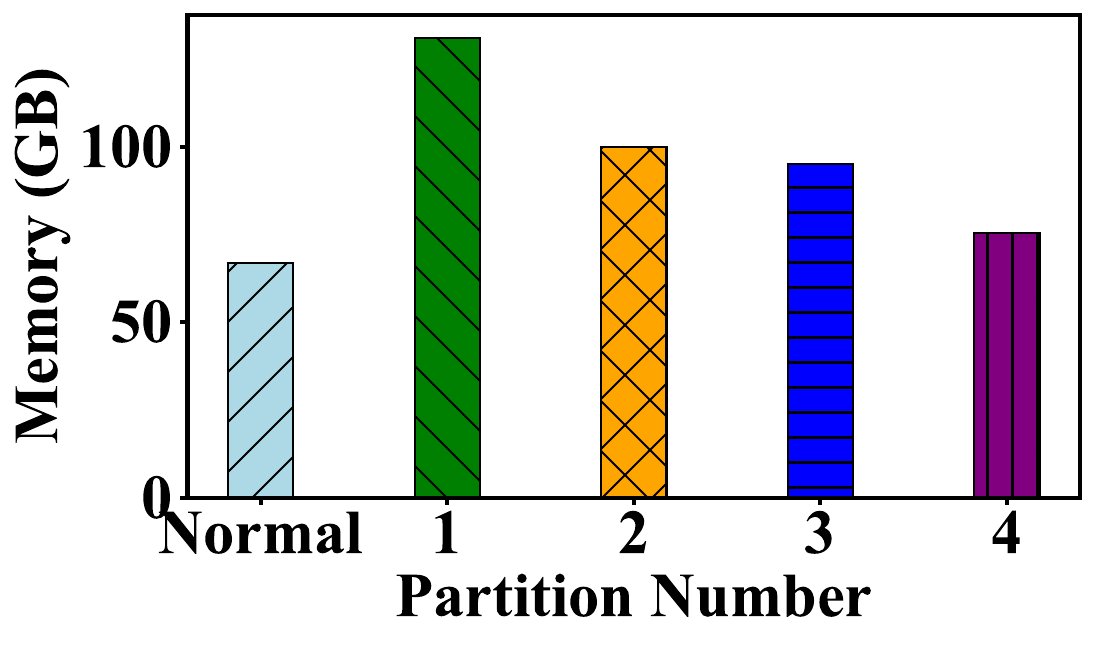}
        \captionsetup{justification=centering}  % Center the caption
        \caption{Total Memory}
        \label{fig:pipeline_partition_performance_d}
    \end{subfigure}

    \caption{Comparison of performance across different partition numbers in the pipeline approach.}
    \label{fig:pipeline_partition_performance}
\end{figure}
\paragraph{Partition number} Last, we evaluate the scalable pipeline framework with fix 4 solve workers against varying partition numbers as shown in \cref{fig:pipeline_partition_performance}. The normal case means no partition and pipeline design.
The results, shown in \cref{fig:pipeline_partition_performance_a}, reveal that the total proof generation time is irrespective of the number of partitions if the memory is affordable, remaining around 160s in all cases with the pipeline. This indicates that partitioning does not introduce significant overhead to the system in terms of time. \cref{fig:pipeline_partition_performance_b} shows that CPU utilization stays consistently high (around 55\%) across different partition configurations, confirming that partitioning does not affect the performance again. Finally, \crefrange{fig:pipeline_partition_performance_c}{fig:pipeline_partition_performance_d} show that increasing the number of partitions leads to substantial reductions in both total memory and prove memory in the scalable pipeline framework. This reduction enables \system to handle multiple larger circuits concurrently without the risk of overwhelming the memory.

\vspace{-5pt}
\section{Related Work}\label{sec:related_work}
\paragraph{\zk protocol and implementation}
A variety of \zk protocols\cite{groth2016size, gabizon2019plonk, chen2023hyperplonk, setty2020spartan, chiesa2020marlin, wahby2018doubly, bunz2018bulletproofs} have been developed to enhance proof efficiency, focusing on reducing proof size and verification time. These innovations have significantly advanced the practicality of zero-knowledge proofs in real-world applications. To further support the practical use of zero-knowledge systems, several frameworks\cite{gnark, libsnark, circom} have been developed, providing tools for both circuit design and proof generation. These frameworks offer high-level APIs for constructing circuits, while handling the underlying cryptographic operations, such as elliptic curve computations and multi-scalar multiplications, ensuring efficient proof generation. Our implementation is based on the Gnark library, a framework that streamlines the development of \zk circuits and the proof generation process.

\paragraph{Optimizations in Proof Computation}  
Many prior works have demonstrated that FFT/NTT computations, an important part of proof computation,  can be grouped and executed in parallel \cite{chen2017big, dai2016cuhe, goey2021accelerating, kim2020accelerating}. This significantly improves computational efficiency and resource utilization. To reduce the complexity of MSM operations in proof computation, approaches such as Pippenger’s \cite{pippenger1976evaluation} and Straus’ \cite{straus1964addition} algorithms have been leveraged, while other works have decomposed MSM into smaller, parallelizable tasks \cite{zhang2021pipezk, ma2023gzkp}. 
With these optimizations, witness generation, in turn, becomes the new bottleneck in the overall process.  Furthermore, Hardware acceleration, especially using GPUs, has also been widely explored \cite{chen2017big, dai2016cuhe, goey2021accelerating, kim2020accelerating, ma2023gzkp, zhang2021pipezk, ji2024accelerating, govindaraju2008high}. These optimizations form a crucial part of improving proof computation performance, and can further be integrated into our framework to enhance the overall system's efficiency.

\paragraph{Memory Reduction}  
Earlier efforts \cite{bootle2022gemini} reduced memory usage by modifying specific protocols, but such approaches were often limited to certain ZKP systems and incurred slower proof generation. Recently, VOLE-based interactive ZKP protocols \cite{baum2021mac, dittmer2020line, weng2021wolverine} have reduced memory requirements, though at the cost of increased bandwidth usage. SPLIT \cite{qi2023split} introduced a novel method for reducing memory bottlenecks by partitioning circuits into smaller subcircuits for sequential execution. However, SPLIT relies on manual circuit partitioning, which limits its scalability and generalization. Instead, our work proposes an automatic partitioning approach that fits more complex circuits.

\paragraph{Distributed \zk}  
Another line of research has focused on distributed ZKP protocols \cite{wu2018dizk, liu2024pianist, xie2022zkbridge, sang2023automating}, which distribute the workload across multiple machines to alleviate memory pressure and enable large-scale proof generation. However, these protocols face limitations, including inter-machine communication overhead and the complexity of coordinating multiple nodes. 
%In scenarios where distributed resources are unavailable or costly, such approaches become impractical. 
We consider distributed \zk as a valuable extension of our work in the future if our optimization for a single machine is still under demand.

%, while our design achieves better efficiency and resource utilization for each individual machine.

\section{Conclusion}\label{sec:conclusion}
\zk is widely used in verifiable outsourcing, yet the computational cost and memory usage during proof generation present significant scalability challenges. Most existing work focuses on optimizing the proof generation phase, largely overlooking the speed and resource utilization issues during the witness generation phase, which remains a significant bottleneck. In this paper, we address these issues by introducing a pipeline architecture and an automatic circuit partitioning algorithm. Our pipeline approach allows for efficient parallelization of witness generation and proof computation, while the circuit partitioning method reduces memory overhead by dividing large circuits into smaller, manageable subcircuits, providing more flexible control over memory usage during proof generation. Experimental results demonstrate that these solutions enable significant performance improvements and resource efficiency.

\bibliographystyle{plain}
\bibliography{references}

\end{document}